\documentclass[epsf]{aastex}
\usepackage{apjfonts}
\usepackage[onecolumn]{emulateapj5}
\usepackage{graphicx} 
\usepackage{color} 
\singlespace
 
\newcommand{\be}{  \begin{eqnarray} }
\newcommand{\ee}{  \end{eqnarray} }

\def\msun{{\rm M}_\odot}

\def\kb{k_{\rm B}}

\def\spose#1{\hbox to 0pt{#1\hss}}
\def\lta{\mathrel{\spose{\lower 3pt\hbox{$\mathchar"218$}}
     \raise 2.0pt\hbox{$\mathchar"13C$}}}
\def\gta{\mathrel{\spose{\lower 3pt\hbox{$\mathchar"218$}}
     \raise 2.0pt\hbox{$\mathchar"13E$}}}
\font\syvec=cmbsy10                        
\font\gkvec=cmmib10                         
\def\bnabla{\hbox{{\syvec\char114}}}       
\def\bxi{\hbox{{\gkvec\char24}}}           

\begin{document}

\title{The Neutrino Bubble Instability:  A Mechanism for Generating
Pulsar Kicks}
\author{Aristotle Socrates,\altaffilmark{1,2} 
Omer Blaes,\altaffilmark{3} Aimee Hungerford\altaffilmark{4}, 
and Chris L. Fryer\altaffilmark{5}}
\altaffiltext{1}{Department of Astrophysical Sciences, Princeton 
University, Peyton Hall-Ivy Lane, Princeton, NJ 08544: 
socrates@astro.princeton.edu\\
and Department of Physics, University of California, Santa Barbara, 
CA 93106}
\altaffiltext{2}{Hubble Fellow}
\altaffiltext{3}{Department of Physics, University of California, 
Santa Barbara, CA 93106: blaes@physics.ucsb.edu}
\altaffiltext{4}{Transport Methods, Los Alamos National Laboratory,
Los Alamos, NM 87545: aimee@lanl.gov\\
and The University of Arizona, Tucson, AZ 85721 }
\altaffiltext{5}{Theoretical Astrophysics, Los Alamos National Laboratory,
Los Alamos, NM 87545: fryer@lanl.gov \\ 
and Physics Department, The University of Arizona, Tucson, AZ 85721}

\begin{abstract}
An explanation for the large random velocities of pulsars
is presented.  Like many other models, we propose that the momentum
imparted to the star is given at birth.  The ultimate source of 
energy is provided by the intense optically thick 
neutrino flux that is responsible for radiating the proto-neutron star's 
gravitational binding energy during the Kelvin-Helmholtz phase.  The central 
feature of the kick 
mechanism is a radiative-driven magnetoacoustic instability, which we refer
to as ``neutrino bubbles.''  Identical in nature to the photon bubble 
instability, the neutrino bubble instability requires the presence
of an equilibrium radiative flux as well as a coherent steady background
magnetic field.  Over regions of 
large magnetic flux densities, the neutrino bubble instability is allowed
to grow on dynamical timescales $\sim 1$ms, potentially leading to 
large luminosity enhancements and density fluctuations.  Local luminosity 
enhancements, which preferentially occur over regions of strong 
magnetic field, lead to a net global asymmetry in the neutrino emission 
and the young neutron star is propelled in the direction opposite to 
these regions.  For favorable values of magnetic field structure, size, 
and strength as well as neutrino bubble saturation amplitude, momentum kicks 
in excess of  $1000\,{\rm km\,s^{-1}}$ can be achieved.  
Since the neutrino-powered kick 
is delivered over the duration of the Kelvin-Helmholtz time $\sim$ a few 
seconds, one expects spin-kick alignment from this neutrino bubble powered
model.  
\end{abstract}

\keywords{stars:evolution--stars:oscillations--MHD--instabilities--
gravitational waves--pulsars:general}

\section{Introduction}

Shortly after their discovery, it was understood that neutron stars travel 
with random velocities significantly in excess of the massive stars from
which they were born (Gunn \& Ostriker 1971).  More recently, studies of 
pulsar velocity distributions suggest that neutron stars indeed possess
large proper motions and that the momentum kick is most likely given at birth
possibly with a bimodal velocity distribution (Lyne \& Lorimer 1994; 
Lorimer, Bailes, \& Harrison 1997; Fryer, Burrows, \& Benz 1998; 
Cordes \& Chernoff 1998; Arzoumanian, Chernoff, \& Cordes 2002).   
The orbital behavior of
individual double neutron star and black hole systems also imply that 
neutron stars receive natal kicks (Fryer \& Kalogera 1997; Kramer 1998; Tauris 
et al. 1999; Wex, Kalogera, \& Kramer 2000; Mirabel et al. 2002). 

From a theoretical point of view, a natal kick is a reasonable outcome
of stellar collapse and the subsequent explosion.  Approximately $10^{51}
\,{\rm erg}$ of kinetic energy is released by the prompt explosion 
mechanism, which can be thought of as as shell of
$\sim 10^{-1}{\msun}$ traveling radially outward at 
$\sim 10^{9}\,{\rm cm\,s^{-1}}$.  Thus, a momentum asymmetry in the shock of 
$\sim 10\%$ could potentially endow a neutron star with a kick velocity 
$\sim 10^{8}\,{\rm cm\,s^{-1}}=10^3\,{\rm km\,s^{-1}}$.  Motivated by this
simple kinematic argument, Burrows \& Hayes (1996) pursued a 
``hydrodynamically-driven'' kick mechanism by artificially deforming the outer 
portions of an Fe core progenitor and numerically evolving the 
subsequent collapse and bounce. They found that a ``mass-rocket'' produced
a kick of $\sim 500\,{\rm km\,s^{-1}}$.  One
possibility for a substantial seed asymmetry may be due to unstable 
{\it g}-modes in progenitor Fe core, which are driven by the 
$\epsilon$-mechanism in the outer shell (Lai \& Goldreich 2000).

The intense optically thick neutrino radiation is another potential 
reservoir of energy for natal kicks.  During the 
Kelvin-Helmholtz phase, neutrino emission is responsible for
 radiating $\sim 99\%$ of 
the star's gravitational binding energy, which amounts to 
$\sim$ a few $\times 10^{53}\,{\rm erg}$.  Asymmetric  
neutrino surface emission of order $\sim$ a few\% leads to a 
kick of $\sim 10^3\,{\rm km\,s^{-1}}$.  The magnetic field of a 
proto-neutron star may provide the needed systematic asymmetry in such a
profoundly spherically symmetric environment, in which the neutrinos can 
couple.  Since the weak interaction violates parity conservation, the 
neutrinos preferentially scatter towards one magnetic pole over the
other.  This ``neutrino/magnetic-driven'' mechanism requires global 
magnetic field strengths $\sim 10^{16}$G in order to produce a kick of 
$\sim 10^{3}\,{\rm km\,s^{-1}}$ (Arras \& Lai 1999a, 1999b).  

In this paper, we describe an alternative model for neutron star kicks 
that can be viewed, in the broadest sense,
as a synthesis of the hydrodynamic- and neutrino/magnetic- driven 
models.  By considering general radiative envelopes that are stratified,
optically thick, magnetized, and highly electrically conducting, 
Blaes \& Socrates (2003, hereafter BS03) showed that 
local magnetoacoustic perturbations are 
susceptible to radiative driving, a phenomenon known as ``photon bubbles''
(Arons 1992; Gammie 1998).  The non-linear evolution of the photon bubble
instability is far from certain as research of this  
phenomenon is in the developmental stages 
(Begelman 2001, Turner et al. 2004).  
Regardless, it is reasonable to suspect that radiative
flux enhancements occur in the saturated state since the radiative flux
provides the free energy for the  instability.  In the case of 
proto-neutron star atmospheres, {\it neutrino bubbles} rather than photon
bubbles may vigorously operate in surface regions that are 
substantially magnetized since neutrinos rather than photons are the 
particle species responsible for the diffusive transfer of energy.   Over
these regions, neutrino flux enhancements propel the star in the opposite
direction, potentially providing a natal kick (See Figure 1).  
In a way, one can 
view this mechanism as a ``neutrino magnetohydrodynamically-driven''
mechanism.  The main focus of this work is to understand the mechanics, 
thermodynamics, and stability properties of local radiation 
magnetohydrodynamic oscillations in proto-neutron stars 
so that we can assess the feasibility, to first approximation, of a 
neutrino bubble powered kick mechanism.  

The organization of this paper is as follows.  In 
\S \ref{s:mechanism} we provide some initial theoretical arguments in order 
 to motivate this work.  The importance of radiation-driven
fluid instabilities is stressed by comparing the envelope structure 
of proto-neutron stars with those on the main sequence.  Based on 
the magnetic field structure of the solar convection zone, we describe 
magnetic field configurations on the surface of proto-neutron stars
that are favorable for neutrino bubble powered kicks.  A detailed 
analysis of local magnetoacoustic perturbations in the limit 
of rapid neutrino diffusion is carried out in \S\S \ref{s:fundamental}-\ref
{s:mechanics}.  Our simplifying assumptions and problem set up are sketched
out in \S\ref{s:fundamental}.  The thermodynamic behavior of short-wavelength
oscillations in degenerate radiating fluids is investigated in 
\S\ref{s:thermodynamics}.  The driving mechanism of the neutrino bubble
instability is presented \S\ref{s:mechanics} in terms of the 
relationship between pressure and density as well as force and velocity.  
Also in \S\ref{s:mechanics}, the neutrino bubble stability criterion 
is evaluated in terms of parameters relevant for proto-neutron star 
surfaces.  A Monte Carlo radiation transfer study is performed in 
\S \ref{s:montecarlo} where we estimate the value of the kick as well as
the gravitational wave strain produced by the globally asymmetric neutrino 
radiation. A comparison to other kick mechanisms and a
critical summary are given in \S \ref{s:conclusions}.

\section{The Kick Mechanism}\label{s:mechanism}

A proto-neutron star (PNS) releases $\sim 99\%$ of the accumulated
binding energy from its Fe core progenitor in the form of neutrinos.
The neutrinos escape the PNS nearly uniformly across all neutrino
flavors during the Kelvin-Helmholtz phase, which lasts for $\sim
10\,{\rm s}$ after its birth.  If the neutrino emission is asymmetric
by roughly a few percent, the PNS receives a natal kick $\sim
10^3\,{\rm km\,s^{-1}}$ (Arras \& Lai 1999a).  Since the neutrino
emission by far dominates the energetics during the Kelvin-Helmholtz
phase, the main difficulty in formulating a neutrino-powered kick
model rests in identifying an intrinsic systematic asymmetry in which
the neutrinos can couple.

A natural candidate for inducing a neutrino asymmetry is the magnetic 
field of the PNS.  On their own, magnetic fields alter neutrino-nucleon 
scattering and absorption via coupling
between the background field 
and the nucleon's magnetic dipole moment.  Since the weak 
interaction violates parity, the introduction of an axial magnetic 
field into the scattering matrix causes the neutrinos to preferentially
scatter toward one magnetic pole over the other.  However, this effect
exists only as long as neutrinos are allowed to deviate from local 
thermal and
chemical equilibrium (Arras \& Lai 1999a).  At and above this layer, 
neutrinos begin to mechanically decouple 
from the matter as well, implying a lack of momentum exchange between
the neutrinos and the entire PNS.
Thus, strong $\sim 10^{16}\,{\rm G}$ global magnetic fields are required
to induce a significant neutrino asymmetry above the energy decoupling
layer.      

Here, we also exploit the magnetic field for the purpose  of generating 
neutrino-powered natal kicks.  Rather than altering the thermal 
transport properties of neutrinos, we propose that the magnetic field 
induces a radiation-driven fluid instability at and below the 
neutrinosphere.  The fluid instability is acoustic in nature 
and is driven in the exact same way as the ``photon bubble'' instabilities,
which may operate in neutron star accretion caps, relativistic accretion 
disks, and magnetized main sequence stellar envelopes (Arons 1992, Gammie 
1998, BS03).  In the case of PNSs, neutrinos
are the species with relatively large mean free paths and therefore 
enable the diffusive transport of energy.  It follows that
in the envelopes of PNSs, the {\it neutrino} bubble instability is the 
relevant mechanism when considering radiative magnetosonic stability.  As 
the instability grows and saturates, a luminosity perturbation
is induced about regions of large magnetic flux density and the PNS is 
propelled in the direction opposite to these ``starspots.''

\begin{figure}[t!]
\begin{center}
\input{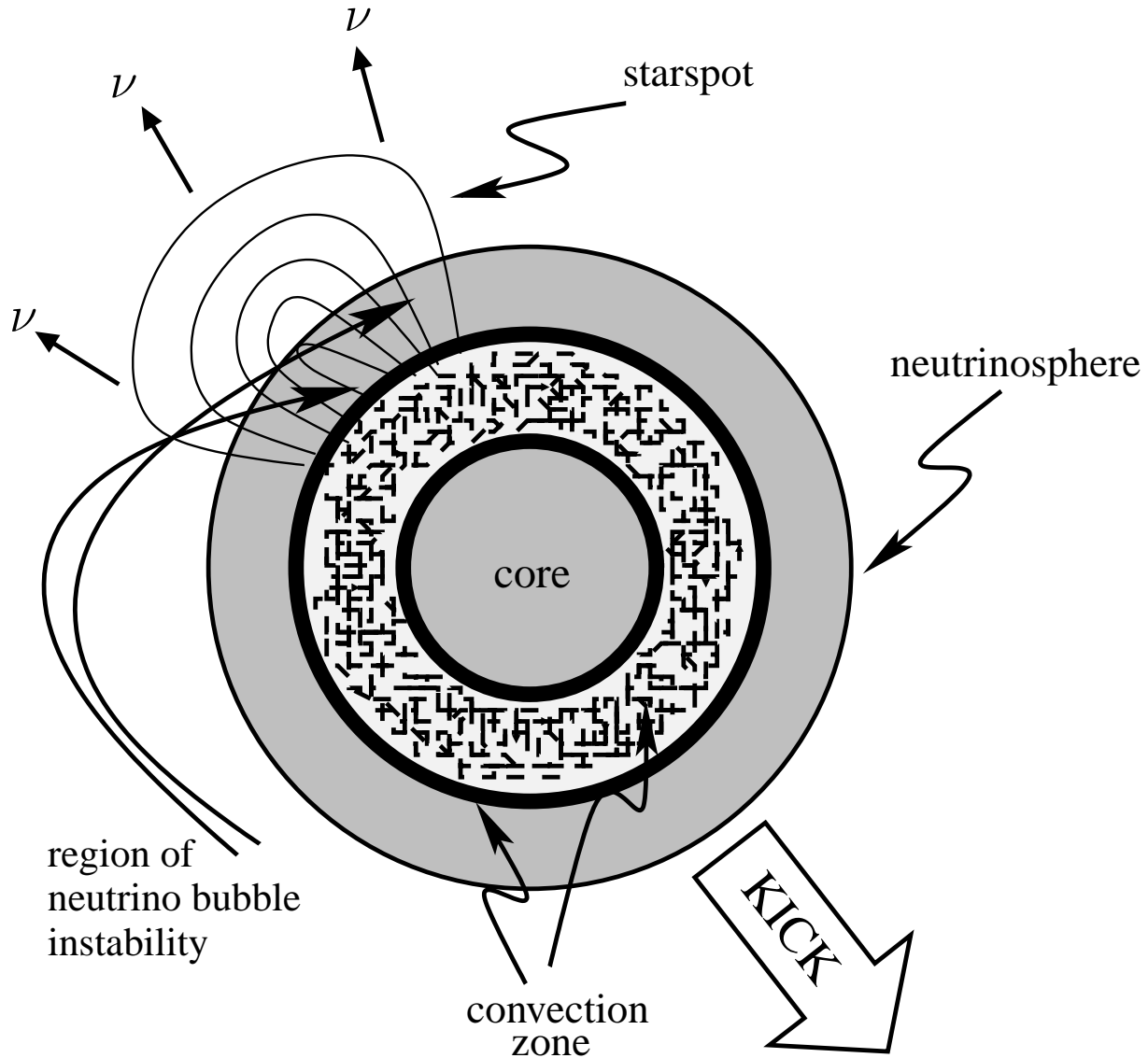}
\caption{Sketch of the kick mechanism.  Starspot-like magnetic
field structures are generated at the base of convection zone and are
transported to the upper-base of the convection zone by both                
buoyancy and turbulent pumping in a few overturn times.  If the presence of 
strong, long-lived, large scale magnetic ``starspots'' can induce 
neutrino bubble instability for a significant fraction of the 
Kelvin-Helmholtz time, the proto-neutron star will receive a natal kick
due to enhanced neutrino transport in the vicinity of the starspot.  
For average initial rotation rates and random starspot placement, 
the rotation axis will most likely be aligned with the kick for a given 
neutron star.}
\end{center}
\label{f:model}
\end{figure}
   
In order for a radiation-driven fluid instability to saturate with a
large luminosity perturbation, one expects the radiation pressure to
be at least comparable in value to the gas pressure.  For PNSs, the
average neutrino luminosity during the Kelvin-Helmholtz era is $\sim
0.01$ the Eddington value, implying that the equilibrium radiation
pressure is $\sim 0.01$ the equilibrium gas pressure.  However, at and
below the neutrinosphere, the radiation and neutrinos are in
thermal and chemical equilibrium with one another on the dynamical
timescales of interest, which are of order 1 ms -- the acoustic
crossing time over a gas pressure scale height.  We will demonstrate
that the spatial distribution of both temperature and neutrino
chemical potential for a given acoustic MHD mode are responsible for
driving the neutrino bubble instability.  Since the the temperature
and relative composition of the gas are locked to that of the
neutrinos, the gas itself behaves in a fashion similar to the
radiation and provides for the majority of the driving.  It follows
that at least to linear order, the stress responsible for radiative
driving is comparable in magnitude to the acoustic restoring force,
opening up the possibility for large saturation amplitudes.

Radiation-driven acoustic instabilities, e.g. the $\kappa$-mechanism, are 
ultimately driven by the background radiation flux coupling to the 
fluid displacement and density perturbation. Yet, damping from radiative 
diffusion only depends on the total pressure of the gas.  The larger 
the ratio of flux to pressure, the more likely driving will overcome 
damping.  In other words, as the radiation field becomes increasingly 
anisotropic, the free energy per quanta available for acoustic driving 
increases accordingly.  Due to their stiff equation of state above
nuclear saturation density and the relatively low number of total 
optical depths $\sim$ a few thousand, the radiation fields of PNSs are highly
anisotropic in comparison to those of main sequence stars whose 
luminosities are $\sim 0.01\,L_{\rm edd}$ in photons.  To illustrate this 
point, compare the fraction of stellar mass lying one optical depth
below the surface of last scattering for a PNS and a 4$M_{\odot}$ A-star
\be
{\rm PROTO-NEUTRON\,\,\,STAR\left(0.01\,L_{\rm edd}\right):}
\,\,\,\,\,\,\,\frac{\delta M}{M_{\odot}}
\sim\frac{R^2_{\nu}\,H_{\nu}\rho_{\nu}}{R^3_{\nu}\rho_c}&=&10^{-4}
\left(\frac{H_{\nu}}{1\,{\rm km}}\right)\left(\frac{10\,{\rm km}}
{R_{\nu}}\right)\left(\frac{\rho_{\nu}}{10^{11}{\rm g\,cm^{-3}}}\right)
\left(\frac{10^{14}\,{\rm g\,cm^{-3}}}{\rho_c}\right)\nonumber\\
{\rm MAIN\,\,\,SEQUENCE\,\,\,STAR\left(0.01\,L_{\rm edd}\right):}
\,\,\,\,\,\,\,\,\frac{\delta M}{M_{\odot}}\sim\frac{R^2_{*}
\,H_{*}\rho_{*}}{R^3_{*}\rho_c}&=&10^{-11}\left(\frac{H_{*}}{10^2\,{\rm km}}
\right)\left(\frac{10^5\,{\rm km}}
{R_{*}}\right)\left(\frac{\rho_{*}}{10^{-7}{\rm g\,cm^{-3}}}\right)
\left(\frac{10\,{\rm g\,cm^{-3}}}{\rho_c}\right)
\ee
where $R_i,H_i, \rho_i,$ and $\rho_c$ is the radius of the surface of 
last scattering, the scale height of the surface of last scattering, 
the density at the last scattering surface, and the central core density,
respectively.  From the above argument, one suspects that a greater
fraction of the total stellar mass is responsible for radiative-acoustic 
driving in PNSs than in main sequence stars of comparable Eddington 
factors.  Thus, PNSs are promising sights for radiation-driven acoustic 
instability of any type since a larger mass fraction responsible for 
driving implies a relatively large luminosity perturbation.  If, for
example, the $\kappa$-mechanism (Baker \& Kippenhanh 1962, Unno et al.
1989) or strange modes (Glatzel 1994) operate in the envelopes of 
PNSs, large luminosity perturbations may follow.  An increase in 
the overall integrated luminosity of $\sim 10\%$ during the first 
few hundred milliseconds may significantly contribute to the explosion
mechanism.  Since the focus of this work is centered on producing a potential 
kick mechanism, our
analysis will be heavily weighted toward radiation-MHD driving 
mechanisms which are only located on patches of large
magnetic flux density.   

The nature of the magnetic field structures in the vicinity of the 
neutrinosphere in part, determines whether or not neutrino bubble 
instability can power kicks, or even explosions.  In the following discussion,
mainly phenomenological arguments are put forth in order to motivate the
{\it possibility} that coherent magnetic formations can exist for a
substantial fraction of the Kelvin-Helmholtz time while covering a sizable
fraction of the star, with large ($10^{14}-10^{15}$G) field strengths.

\subsection{Convection and Magnetism in Proto-Neutron Stars}
\label{ss:magconv}

It is believed that PNSs convect as they cool.  At very early times
the escaping shock dissociates ambient nuclei into baryons, heating
the mantle while producing a negative entropy gradient that drives
convection (Bethe, Brown \& Cooperstein 1987; Wilson \& Mayle 1988).
Due to their large energies and mean free paths, neutrinos radiate the
entropy excess left over by the prompt shock, immediately quenching
the convection in $\lesssim 100$ ms.  Soon after this somewhat
superficial episode of convective overturn, a PNS suffers from deep
convective instability for the duration of the Kelvin-Helmholtz time
(Thompson \& Murray 2001, hereafter TM01; Burrows \& Lattimer 1986
Burrows 1987; TD93; Keil, Janka, \& Mueller 1996, hereafter KJM96;
Pons et al. 1999).  KJM96 find that Ledoux-type convection quickly
recedes from the neutrinosphere in $\sim 20-30$ ms, moving deep into
the core and at about $\sim 1$s after bounce, the entire star is
convective.  TD93 point out that in PNSs, the ratio of convective
kinetic energy to gravitational binding energy $\sim 100$ times larger
than its progenitor during the Silicon burning phase and $\sim
10^3-10^4$ times larger when the progenitor resided on the main
sequence.  In fact, when compared this way, convection in PNSs is
$\sim 10^4$ times more violent than the Sun.  Primarily motivated by
this fact, TD93 conclude that PNSs are efficient sites of dynamo
activity.

The theory of magnetic field generation and its subsequent evolution in 
young PNSs is primarily put forth 
by TD93 and then further elaborated by TM01.  A summary of their 
findings, relevant for this work, is given below.    

\subsubsection{Case of Rapid Rotation--Magnetars}

Based on theoretical and 
phenomenological arguments, Duncan and Thompson (1992) and TD93 conclude that
PNSs with initial roation periods near break up ($P_{\rm rot}\sim 1$ ms)
efficiently convert the nascent star's rotational
shear energy into toroidal and poloidal flux via $\alpha-\Omega$ dynamo 
action at the base of the convection zone where the overturn time is 
similarly $\sim 1$ ms. They argue that the free energy available from 
differential rotation $E_{\Omega}$ for rapidly rotating 
($P_{\rm rot}\sim 1$ ms)
PNSs is roughly 
$E_{\Omega}\sim 10^{52}\,{\rm ergs}$.  This enormous reservoir of energy could 
in principle, generate ordered fields
as strong as $B\sim 3\times 10^{17}\,{\rm G}$ throughout the star.  This led 
them to
conclude that PNSs born rotating near break up form dipole fields
during the convective Kelvin-Helmholtz phase that are much stronger
than those of pulsars.  These highly magnetized neutron stars with dipole 
field strengths $B_{\rm dip}\sim 10^{14}-10^{15}\,{\rm G}$ are 
referred to as ``magnetars.''  

\subsubsection{Slow Rotation -- Analogy with the Solar Convection Zone}

In the absence of rapid rotation or a large reservoir of helicity,
turbulent convection alone may lead to sizable magnetic stresses.
There is some phenomenological evidence for this idea, most of which
comes from the Sun.  Between 1986 and 1988, a period of increasing
solar magnetic activity, the solar 5-minute acoustic oscillations
displayed a significant (one part in $10^{4}$) increase in frequency
(Libbrecht \& Woodard 1990).  A rise in the
rms magnetic stress to values of $B_{\rm rms}\sim 200\,$G at the top of the
convection zone, and extending downwards to substantial depth with 
greater strength, could explain the observed trend in acoustic frequency (Goldreich
et. al 1991).  At the top of the solar convection zone, the overturn
time ($\sim 5-10$ min) is much shorter than the rotation period ($\sim
22$ days).  Therefore, it seems that convective motions alone may sustain 
near equipartition (with the turbulence) rms fields on timescales that 
are many orders of magnitude in excess of the convective overturn time.

In PNSs,  unlike the solar convection zone, the gradient of the 
turbulent diffusivity lies in 
the opposite direction with respect to the buoyancy force acting on a 
given flux rope in thermal equilibrium with its surrounding (Tobias et. al 
1998, 2001; TM01).  As a result, both buoyancy and turbulent diffusion 
tend to transport magnetic structures of scale $l_P$, the gas pressure scale
height, to the top of 
the convection zone approximately in an overturn time.\footnote{This is true 
so long as magnetic flux is {\it passively} mixed by the turbulence.}  At the
same time, flux ropes that are generated and amplified by the stretching
action of the convective cells are pinned at the base of the overshoot 
region.   

One of the most commonly held beliefs in 
dynamo theory is that turbulent magnetic fields generated and amplified 
by convective overturn come roughly into equipartition with the turbulent
kinetic energy.  Mixing length theory provides for a rough estimate of the 
convective velocities at the base of the convection zone
\be
F_c\sim \rho\,v_c^3\longrightarrow  v_c\sim 2\times10^{8}\,F^{1/3}_{39}
\,\rho^{-1/3}_{14}\,{\rm cm\,s^{-1}}
\label{MLT}
\ee
where $F_c$ and $v_c$ are the convective energy flux and velocity 
amplitude, respectively.  This yields a saturation field strength 
\be
B_{\rm sat}\sim \sqrt{4\pi\rho}v_c\sim 6\times 10^{15}\,F^{1/3}_{39}
\,\rho^{1/6}_{14} {\rm G}
\label{Bsat}
\ee
on the scale of a pressure scale height $l_P$ at the base of the convection
zone where $l_P\sim 1-10\,{\rm km}$.  If a near-equipartition 
magnetic flux tube generated at 
the the base of the convection zone is ``pumped'' and pinned to the top 
of the PNS convection zone in the overshoot region (KJM96, TM01), then the 
local field strength may approach equipartition {\it with the matter} near
the neutrinosphere for typical values of temperature and 
density ($T_{\nu}\sim$ a few MeV and $\rho_{\nu}\lesssim 10^{12}\,{\rm g\,
cm^{-3}}$) at relatively small depths.

\subsubsection{Long-Lived Large Scale Magnetic Structures in the 
Limit of Slow Rotation}

The ratio of the length scale of the energy bearing eddies to the radius of 
the neutrinosphere, $l_p/R_{\nu}\sim 10^{-1}$, is $\sim 10^3$ times 
larger than the analogous ratio for the Sun.  This leads one 
to believe that the magnetic 
structures that are produced by convective turbulence in PNSs are much larger
than those of the Sun as a fraction of the total stellar radius (TD93).  
The value of the fluctuating rms magnetic field is given by
\be
B_{\rm rms}\sim \epsilon^{1/2}_B\,B_{\rm sat},
\ee 
where $\epsilon_B$ is the convective efficiency parameter.  In the Sun, 
$\epsilon_B\sim 0.1$ (Goldreich et al. 1991) and if we naively 
extrapolate this value to PNSs, we
obtain with the help of eq. (\ref{Bsat}) and with all things being equal, 
$B_{\rm rms}\sim 2\times 10^{15}$.  One suspects that these magnetic 
structures, with spatial dimension roughly equal to the depth of the
radiative layer above the convection zone, are created and destroyed on the 
eddy turnover time $\sim 1$ms.  However, Cattaneo and Vainshtein (1991) point 
out that larger scale structures with smaller flux densities can effectively 
resist eddy diffusivity for a significant number of overturns, implying 
a breakdown of kinetic approximation.  
If a starspot possesses a dimension $l_{\rm spot}\sim 3 l_p$ and is due to
the random orientation of the smaller dipoles created on the scale of the 
energy bearing eddies, its field strength $B_{\rm spot}\sim 1/10B_{\rm rms}
\sim 2\,(\epsilon_B/0.1)^{1/2}\times 10^{14}\,{\rm G}$.  Such a starspot 
covers $\sim$ a few $\%$ of the surface of the PNS and may exist for 
several hundred turnover times before eddy diffusivity can efficiently 
transfer flux from large scale structures to the eddy 
scale (Cattaneo \& Vainshtein 1991). 

From an even more phenomenological perspective, supergranulation on the
surface of the Sun may provide a clue to the mechanism of large scale
field generation in slowly rotating PNSs.  In the Sun's quiet
photospheric network, magnetic flux is generated and destroyed by
granule and supergranule motions, rather than originating from the
strongly magnetized active regions (Schrijver et al. 1997).  The
intermittent magnetic fields associated with supergranule motion
typically approach equipartition with the supergranule flow while
lasting only a few supergranule lifetimes.

The characteristic scale of a solar supergranule is $\sim 30$ times larger 
than the gas pressure scale height at the photosphere or equivalently, 
$\sim 30$ times larger than a characteristic granule (Stenflo 1989).  Also, 
supergranules overturn with velocities
that are $\sim$ a few times slower than the velocities associated with 
granular motion, which are $\sim 1\,{\rm km\,s^{-1}}$ in the Sun.  If 
we savagely
map the solar values for granule and supergranule motion to the surfaces of 
PNSs, one expects supergranule lifetimes on the surface of the PNSs
to be anywhere from a few hundreds of milliseconds to a few seconds.  The
scale height of the neutrinosphere $\sim 1/30$ the stellar radius 
measured at the neutrinosphere $R_{\nu}$, implying that only a single 
supergranule may encompass a sizable fraction of a PNS.        

So far, we have provided the impetus for studying neutrino bubble 
instabilities in PNSs by motivating the existence of a small number of 
strongly magnetized starspots. Now, we consider in detail the 
local dynamics and 
thermodynamics of their linear driving mechanism.  The next few sections 
are quite technical.  They are riddled with large algebraic 
expressions whose meanings may not be immediately transparent to the 
untrained eye.  For readers not concerned with the  
details of linear radiation magnetohydrodynamics in degenerate media, 
we suggest skipping ahead to \S\ref{sec:stability} where estimates of 
stability and driving are given in terms of physical parameters 
appropriate for PNSs.

\section{Fundamental Assumptions and Equilibrium}\label{s:fundamental}

We consider fluid perturbations upon a highly conducting magnetized 
fluid consisting of 
neutrons, protons, electrons, and photons interacting with radiation
consisting of neutrinos (and anti-neutrinos) of all flavors. The interaction
between normal matter and the neutrinos is mediated by the various sources
of opacity.  We make the simplifying assumption that all neutrino opacities 
increase with the square of the neutrino energy $E$.  In particular, for 
electron neutrinos and anti-neutrinos,
\begin{equation}
\kappa_{\nu_e,\bar{\nu_e}}=\kappa_0\left(E\over E_0\right)^2,
\label{opacity}
\end{equation}
whereas for all other types of neutrinos,
\begin{equation}
\kappa_{\nu_X}=\kappa_{0X}\left(E\over E_0\right)^2.
\end{equation}
(We adopt the same constant fiducial energy scale $E_0$ in both cases.)
The prefactors $\kappa_0$ and $\kappa_{0X}$ are assumed to be constant
and independent of composition.  Only the electron-type neutrinos can be 
absorbed and emitted by pair capture processes
\be
e^-+p\leftrightarrows n+\nu_e\nonumber\\
e^++n\leftrightarrows p+{\bar{\nu_e}}
\ee
implying that only the electron-type neutrinos can exchange energy with 
the matter while all neutrinos can exchange momentum with the matter
via elastic neutrino-nucleon scattering
\be
\nu+N\leftrightarrows \nu+N
\ee 
where $\nu$ represents any of the six ($\nu_e,{\bar\nu_e},\nu_X, {\bar
\nu_X}$) neutrino species and $N$ denotes either neutrons or protons.
                                                                                 
For now, we shall assume that thermal equilibrium 
\be
T_g=T_{\nu},
\ee
where $T_g$ is the gas temperature of the gas, and chemical equilibrium
\be
\mu_p+\mu_e=\mu_n+\mu_{\nu_e},
\ee
where $\mu_i$ is the chemical potential of the $i^{\rm th}$ species, hold 
in both the equilibrium {\it and} the perturbed state\footnote{At temperature
and densities typical for young PNSs, strong and electromagnetic 
interactions easily maintain thermal balance between 
baryons, electrons, and photons.  Also since these species are so well 
coupled to one another, their mean free paths are vanishingly 
small.  For these reasons, from here on we shall refer 
to ``the fluid'' or ``the gas''
as the multi-species gas consisting of neutrons, protons, electrons, and 
photons.}.  This assertion is more subtle than it appears at first glance 
and we discuss its accuracy later on in this work.  

Assuming that the neutrinos are optically thick and that the neutron-
proton-electron-photons ({\it npe}$\gamma$) fluid is an excellent conductor, 
the expressions for mass, momentum, lepton fraction, energy, and 
magnetic flux conservation are                            

\begin{equation}
{\partial\rho\over\partial t} +\bnabla\cdot\left(\rho{\bf v}\right)=0,
\label{eqcont}
\end{equation}
\begin{equation}
\rho\left({\partial{\bf v}\over\partial t}+{\bf v\cdot\bnabla v}\right)
=-{\bnabla}P+\rho\,{\bf g}+{1\over4\pi}({\bf\bnabla\times B}){\bf\times B},
\label{gasmom}
\end{equation}
\begin{equation}
n\left({\partial Y_L\over\partial t}+{\bf v\cdot\bnabla}Y_L\right)
=-{\bf \bnabla\cdot F}_L,
\label{eqlepton}
\end{equation}
\begin{equation}
{\partial U\over\partial t}+{\bf v\cdot\bnabla}U+(U+P){\bf \bnabla\cdot v}
= -{\bf \bnabla\cdot F},
\label{eqenergy}
\end{equation}
\begin{equation}
{\partial{\bf B}\over\partial t}={\bnabla\times}({\bf v\times B}),
\label{dbdt}
\end{equation}
and
\be
\bnabla\cdot{\bf B}=0
\ee
respectively.  Here, $\rho,\,{\bf v},\,P,\,{\bf g},\,{\bf B}
,\,U,\,{\bf F},\,Y_L,\,{\rm and}\,\,{\bf F}_L$ are the fluid 
density, fluid momentum, total ({\it npe}$\gamma$+neutrinos) pressure, 
local tidal gravitational field, the magnetic field,
total energy density, neutrino flux, lepton fraction, and lepton 
flux, respectively.    

Since PNSs are optically thick to neutrinos of all types, we assume
that in eqs. (\ref{gasmom}), (\ref{eqlepton}), and (\ref{eqenergy})
the diffusion approximation accurately describes energy and 
lepton transport
as well as momentum exchange.  Thus, we only consider fluid 
perturbations that are optically thick to neutrino diffusion, restricting
our analysis to the region below the surface of last scattering i.e., 
below the
neutrinosphere.  The radiative energy flux ${\bf F}$ is 
given by the sum
\begin{equation}
{\bf F}={\bf F}_{\nu_e}+{\bf F}_{\bar{\nu_e}}+{\bf F}_{\nu_X}
\label{Ftotal}
\end{equation}
where ${\bf F}_i$ is the energy flux of the $i^{\rm th}$species.  The 
energy flux for electron-type and $X-$type neutrinos is given by
\begin{equation}
{\bf F}_{\nu_e}+{\bf F}_{\bar{\nu_e}}=-{E_0^2\kb^2\over6\pi^2\kappa_0\rho
\hbar^3c^2}{\bnabla}\left({T^2\eta_{\nu_e}^2\over2}+
{\pi^2\over6}T^2\right)
\label{Fnuetotal}
\end{equation}
and
\begin{equation}
{\bf F}_{\nu_X}=-{E_0^2\kb^2\over18\kappa_{0X}\rho
\hbar^3c^2}{\bnabla}T^2,
\label{Fnux}
\end{equation}
respectively.  Since $X-$type neutrinos are created in the neutral current 
pair annihilation process, an excess of $\mu$- or $\tau$-type neutrinos 
is absent.  As a result, only electron-type neutrinos carry lepton number 
due to their ability of being absorbed and emitted during pair capture
and emission processes.  The lepton flux is given by   
\begin{equation}
{\bf F}_L=-{E_0^2\,k_B\over6\pi^2\kappa_0\rho\hbar^3c^2}
{\bnabla}\left(T\,\eta_{\nu_e}\right).
\label{Flepton}
\end{equation}
The expressions for neutrino energy and lepton fluxes as well 
as other important thermodynamic quantities obtained from Fermi-Dirac
distributions are briefly derived in Appendix \ref{a:a}.  The degeneracy parameter
$\eta_i=\mu_i/k_BT$ will be interchanged with $\mu_i$ throughout this 
work depending on the particular application involved.  The mass density 
is given in terms of the nucleon or baryon number density
\begin{equation}
\rho=nm_n,
\end{equation}
where $m_n$ is the mass of the neutron, implying that the
mass difference between the neutron and proton has been ignored
when evaluating the density.  The total pressure and energy density are 
given by
\begin{equation}
P={1\over3}(U_{e^-}+U_{e^+}+U_\gamma+U_{\nu_e}+U_{\bar{\nu_e}}+U_{\nu_X})
+\frac{2}{3}U_b
\end{equation}
and
\begin{equation}
U=U_{e^-}+U_{e^+}+U_\gamma+U_{\nu_e}+U_{\bar{\nu_e}}+U_{\nu_X}+U_b
\end{equation}
where
\begin{equation}
U_b=U_b(\rho,T,Y_p,Y_n),
\end{equation}
\begin{equation}
U_{e^-}+U_{e^+}={7\pi^2\kb^4 T^4\over60\hbar^3c^3}\left[1+{30\over7}
\left({\eta_e\over\pi}\right)^2+{15\over7}\left({\eta_e\over\pi}\right)^4
\right],
\end{equation}
\begin{equation}
U_\gamma={\pi^2\kb^4T^4\over15\hbar^3c^3}=aT^4,
\end{equation}
\begin{equation}
U_{\nu_e}+U_{\bar{\nu_e}}={7\pi^2\kb^4 T^4\over120\hbar^3c^3}
\left[1+{30\over7}\left({\eta_{\nu_e}
\over\pi}\right)^2+{15\over7}\left({\eta_{\nu_e}\over\pi}\right)^4\right],
\end{equation}
\begin{equation}
\,\,\,\,\,\,{\rm and}\,\,\,\,\,\,U_{\nu_X}+U_{{\bar\nu}_X}=
4\times\frac{7}{8}aT^4.
\end{equation}
The electron and electron neutrino fractions are given by
\be
Y_e=\frac{n_{e^-}-n_{e^+}}{n}={\kb^3T^3\over3n\hbar^3c^3}
\eta_e\left(1+{\eta_e^2\over\pi^2}\right)
\,\,\,\,\,\,\,\,{\rm and}\,\,\,\,\,\,\,Y_{\nu_e}=
\frac{n_{\nu_e}-n_{{\bar\nu_e}}}{n}={\kb^3T^3\over6n\hbar^3c^3}
\eta_{\nu_e}\left(1+{\eta_{\nu_e}^2\over\pi^2}\right),
\ee 
where the total lepton fraction is defined as $Y_L=Y_e+Y_{\nu_e}$.  The 
baryon energy density, $U_b$, is given by that of a monatomic ideal 
gas near the neutrinosphere, where the effects of degeneracy pressure
are negligible for the baryons.  However, in the core the zero-point
Fermi pressure is substantial for the baryons, which is why we have 
written the baryon equation of state in a somewhat general manner since
its form depends on the location below the neutrinosphere.

The condition for chemical equilibrium may be written as
\be
{(m_n-m_p)c^2\over\kb T}-\ln\left[\left({m_n\over m_p}\right)^{3/2}\left({Y_p
\over Y_n}\right)\right]=\eta_e-\eta_{\nu_e}
\label{beta}
\ee
and the fact that the total nucleon number density $n$ is conserved 
implies
\begin{equation}
Y_p+Y_n=1.
\label{baryon}
\end{equation}
Finally, charge neutrality implies
\begin{equation}
Y_p=Y_e={\kb^3T^3\over3n\hbar^3c^3}\eta_e\left(1+
{\eta_e^2\over\pi^2}\right).
\label{charge}
\end{equation}
                                                                                 This gives us 21 equations for 21 variables:  $\rho$, ${\bf v}$, $P$,
${\bf B}$, $U$, ${\bf F}$, $n$, $Y_L$, $T$, $\eta_e$, $\eta_{\nu_e}$,
$Y_p$, $Y_n$,
$({\bf F}_{\nu_e}+{\bf F}_{\bar{\nu_e}})$, ${\bf F}_{\nu_X}$, 
${\bf F}_L$,$U_b,$
$(U_{e^-}+U_{e^+})$, $U_\gamma$, $(U_{\nu_e}+U_{\bar{\nu_e}})$, and
$U_{\nu_X}$.  

We treat the background state of the PNS envelope in a fashion similar
to that of a normal main sequence star.  The gravitational field is assumed 
to be fixed,
which is a good approximation for our oscillation wavelengths of interest 
near the  neutrinosphere, where the 
radiative flux is roughly constant with depth as well.  That is, we adopt a
plane parallel approximation.  Rather than photons
transporting energy from one depth to the next, neutrinos of all 
flavors take on that role.  Also, we implicitly assume that chemical and 
thermal equilibrium holds for our static background and perturbations.  
For PNSs, this condition is met for any dynamical timescale of interest 
below the neutrinosphere (see Appendix \ref{a:c} for a detailed discussion 
regarding chemical and thermal equilibrium).    

The major qualitative difference between the environments of 
PNSs and the stellar envelopes considered by Blaes \& Socrates (2003)
is that the radiation considered here is degenerate with a large 
chemical potential.  Since the flow is optically thick, gradients in 
both neutrino temperature and chemical potential transport both
neutrino energy (not just heat) and particle number. For the 
black body ($\mu=0$) radiation fields considered by Blaes \& Socrates
(2003), 
the background radiation flux interacts with short-wavelength
compressible fluid perturbations in two ways.  Radiation piles up 
downstream density maxima with respect to the background flux ${\bf F}$, 
an effect known 
as ``shadowing'' (Lucy \& White 1980, Lucy 1982).  Thus, the
 radiation flux couples
to the density perturbation, producing a pressure response
$\propto {\bf F}\cdot\bnabla\delta\rho$.  The other effect arises from
changes in a fluid element's position during oscillation with respect to 
surfaces of constant background radiation pressure, which results in 
a pressure response $\propto{\bxi}\cdot{\bf F}$ where $\bxi$ is the 
fluid displacement.  In the case of PNSs, both ${\bf F}$ and ${\bf F}_L$ 
interact with compressible short wave-length fluid oscillations in 
this manner.  Therefore, neutrino bubble instabilities are driven by 
both ${\bf F}$ and ${\bf F}_L$ rather than just ${\bf F}$ alone.  
More fundamentally, radiative diffusion due to gradients in both 
temperature and chemical potential are responsible for radiative 
driving, but otherwise the mechanics of photon bubble-like
instability is the same as the neutrino bubble instability.

\section{Total Pressure Perturbation in the Limit of Rapid Neutrino 
Diffusion}\label{s:thermodynamics}

We wish to study standard magnetosonic motion altered by the presence 
of radiative diffusion.  In the optically
thick limit, radiation alters the fluid's motion by gradients 
in its pressure.  The instabilities classified by Blaes \& 
Socrates (2003) occur in the limit where radiation diffuses 
rapidly compared to the local magnetoacoustic crossing time.  We consider 
the same limit here.

Generally speaking, the total pressure may be written as
\be
P=P\left(\rho,\,T,\,\mu_e,\,\mu_{\nu_e}\right)
\ee
and upon perturbation, we have
\be
\delta P=\delta\rho \left(\frac{\partial P}{\partial\rho}\right)
_{T,\mu_{e},\mu_{
\nu_e}}+\delta T\left(\frac{\partial P}{\partial T}\right)_
{\rho,\mu_{e},\mu_{\nu_e}} +\delta\mu_e\left(\frac{\partial P}
{\partial \mu_e}\right)_{\rho,T,\mu_{\nu_e}}+\delta\mu_{\nu_e}
\left(\frac{\partial P}{\partial\mu_{\nu_e}}\right)_{\rho,T,\mu_e}.
\ee
We may eliminate $\mu_e$ in terms of $\rho, T,$ and $\mu
_{\nu_e}$ by use of the eqs. (\ref{beta}), (\ref{baryon}),
and (\ref{charge}) which denote beta equilibrium, baryon 
conservation, and charge conservation, respectively.  The total
pressure perturbation may then be written as 
\be
\delta P=\delta\rho \left(\frac{\partial P}{\partial\rho}\right)
_{T,\mu_{
\nu_e}}+\delta T\left(\frac{\partial P}{\partial T}\right)_
{\rho,\mu_{\nu_e}} +\delta\mu_{\nu_e}
\left(\frac{\partial P}{\partial\mu_{\nu_e}}\right)_{\rho,T}.
\label{deltaP}
\ee

As previously mentioned, neutrinos diffuse down gradients of both 
temperature and neutrino 
chemical potential.  In the limit of rapid diffusion, the spatial 
distribution and relative amplitude of the neutrino temperature 
and chemical potential are almost purely determined by the
divergence of the neutrino and lepton flux.  To see this clearly,
we write the linearized expressions for conservation of
lepton number, eq. (\ref{eqlepton}), and the first law of thermodynamics,
eq. (\ref{eqenergy}).  This gives us 

\be
-i\omega\delta Y_L+\delta{\bf v}\cdot{\mathbf\nabla} Y_L
=-i\omega\left[\left(\frac{\partial Y_L}
{\partial T}\right)_{\rho,\mu_{\nu_e}}
\delta T+\left(\frac{\partial Y_L}{\partial 
\mu_{\nu_e}}\right)_{\rho, T}
\delta\mu_{\nu_e}+ \left(\frac{\partial Y_L}
{\partial\rho}\right)_{T,\mu_{\nu_e}}
\delta\rho   \right]+\delta{\bf v}\cdot{\mathbf\nabla}Y_L
=-\frac{i}{n}{\bf k}\cdot\delta{\bf F_L}
\label{deltaYL}
\ee
\be
-i\omega\delta U+\delta{\bf v}\cdot {\mathbf\nabla} U+
i\,(U+P)\,{\bf k}\cdot\delta{\bf v}&=&
-i\omega\left[\left(\frac{\partial U}
{\partial T}\right)_{\rho,\mu_{\nu_e}}
\delta T
+\left(\frac{\partial U}{\partial\mu_{\nu_e}}\right)_{\rho,T}
\delta\mu_{\nu_e}+ \left(\frac{\partial U}{\partial\rho}\right)_{T,
\mu_{\nu_e}}\delta\rho\right]\nonumber\\
&+&\delta{\bf v}\cdot {\mathbf\nabla} U+
i\,(U+P)\,{\bf k}\cdot\delta{\bf v}=-i{\bf k}\cdot\delta{\bf F}.
\label{deltaenergy}
\ee
$\delta{\bf F_L}$ and $\delta{\bf F}$ are given by 
\be
\delta{\bf F_L}=-\frac{\delta\rho}{\rho}{\bf F_L}-
\frac{i{\mathcal K}}{\pi^2\kappa_0\rho}{\bf k}
\,\delta\mu_{\nu_e}
\label{deltaFL}
\ee
and
\be
\delta{\bf F}=
-\frac{\delta\rho}{\rho}{\bf F}-\frac{i{\mathcal K}
k^2_BT}{\kappa_{\rm eff}\rho}\,{\bf k}\,\,\delta T-
\frac{i{\mathcal K}\mu_{\nu_e}}{\pi^2\kappa_0
\rho}{\bf k}\,\delta\mu_{\nu_e}
\label{deltaF}
\ee
where 
\be
{\mathcal K}\equiv \frac{E^2_0}{6\hbar^3c^2}\,\,\,\,\,\,{\rm and}
\,\,\,\,\,\,\frac{1}{\kappa_{\rm eff}}\equiv \frac{2}{3\kappa_{0X}}
+\frac{1}{3\kappa_0}.
\ee 
Note that we are operating within the framework of the WKB approximation 
where all perturbation quantities take on a plane wave-like form
$\propto e^{i\left({\bf k}\cdot{\bf x}-\omega t\right)}$.  Therefore, our 
analysis from here on, is only credible for perturbations
with wavelengths smaller than the gas pressure scale height.

From eq. (\ref{deltaFL}), the change in lepton
flux is determined by gradients in the 
neutrino chemical potential mediated by the electron neutrino 
opacity as well as a component that arises from changes in density.  
Eq. (\ref{deltaF}) implies that changes in the 
radiative heat flux originate from both changes in the 
temperature gradient mediated by the sum of the electron- and 
$X$-type neutrino opacities and gradients in the neutrino 
chemical potential mediated solely by the electron neutrino
opacity in addition to a component that comes from fluctuations in 
density.         

A first step in obtaining the total pressure perturbation
is to solve for $\delta\mu_{\nu_e}$ in terms of the other 
thermodynamic variables, $\delta T$ and $\delta\rho$.  We
(arbitrarily) choose the linearized expression for lepton 
conservation, eq. (\ref{deltaYL}) as our starting point.
We have,
\be
-i\,n\,\omega\left[\left(\frac{\partial Y_L}
{\partial T}\right)_{\mu_{\nu_e},\rho}
\delta T+\left(\frac{\partial Y_L}{\partial 
\mu_{\nu_e}}\right)_{T,\rho}
\delta\mu_{\nu_e}+ \left(\frac{\partial Y_L}
{\partial\rho}\right)_{T,\mu_{\nu_e}}
\delta\rho   \right]+n\,\delta{\bf v}\cdot{\mathbf\nabla}Y_L
=i\frac{\delta\rho}{\rho}\,{\bf k}\cdot{\bf F_L}-
\frac{k^2\,{\mathcal K}}{\pi^2\kappa_0\rho}\,
\delta\mu_{\nu_e}.
\ee
It is useful to define a characteristic diffusion frequency,
$\omega_{\rm L,diff}$ 
associated with $Y_L$ transport
\be
\omega_{\rm L,diff}\equiv \omega+i\frac{k^2{\mathcal K}}
{\pi^2\,\kappa_0\,\rho\,n\left(\partial Y_L/\partial\mu_{\nu_e}\right)
_{\rho,T}},
\ee 
which allows us to write
\be
\delta\mu_{\nu_e}=-\frac{\omega}{\omega_{\rm L,diff}}
\frac{\left({\partial Y_L}/{\partial\rho}\right)_{T,\mu_{\nu_e}}}
{\left({\partial Y_L}/{\partial\mu_{\nu_e}}\right)_{\rho,T}}
\,\delta\rho-\frac{{\bf k}\cdot{\bf F_L}}{n\,\omega_{\rm L,diff}
\left({\partial Y_L}/{\partial\mu_{\nu_e}}\right)_{\rho,T}}
\frac{\delta\rho}{\rho}-\frac{\omega}{\omega_{\rm L,diff}}
\frac{\left({\partial Y_L}/{\partial T}\right)_{\rho,\mu_{\nu_e}}}
{\left({\partial Y_L}/{\partial\mu_{\nu_e}}\right)_{\rho,T}}\,
\delta T
-i\frac{\delta{\bf v}\cdot\bnabla Y_L}
{\omega_{\rm L,diff}\left({\partial Y_L}
/{\partial\mu_{\nu_e}}\right)_{\rho,T}}.
\label{deltaetalong}
\ee

BS03 found that in the limit of rapid radiative diffusion, 
temperature
perturbations become reduced on the scale of the mode 
wavelength as long as the characteristic diffusion timescale is 
much shorter than the mode's period.  Here, both the 
 temperature {\it and} neutrino chemical potential are subject to 
the same diffusive reduction since neutrino diffusion occurs
along gradients of both temperature and chemical potential.
Following BS03, we recognize that in the rapidly diffusing limit,
\be
\delta\mu_{\nu_e}\,,\,\delta T\propto\frac{\omega}{\omega_
{\rm diff}}\delta\rho\propto k^{-1}\delta\rho
\ee 
where $\omega_{\rm diff}$ represents some characteristic 
diffusion frequency.  Note that we have assumed that $\omega\propto k$, 
which implies that we are investigating the properties of short-wavelength
MHD waves i.e., the Alfv\'en, fast, and slow waves.  
Many authors have noted that in general, there are  
two diffusion frequencies associated for both lepton and 
energy transport in optically thick PNSs (Bruenn \& Dineva 1996,
Pons et al. 1999, Miralles et al. 2000, Bruenn, Raley, \& 
Mezzacappa 2004).  That is, lepton and energy diffusion together evolve 
down gradients of both lepton fraction and entropy (or alternatively, 
temperature and neutrino chemical potential).  Rather than 
investigating overturn instability phenomena such as Ledoux convection, 
``neutron-fingers,'' and semi-convection, we focus on questions of 
acoustic stability in the limit where sound waves (and magnetosonic waves)
are highly non-adiabatic.  That is, both lepton and energy diffusion occur 
rapidly in comparison to the acoustic crossing time of 
interest for a given oscillation wavelength.  Due to our $\propto E^2$
opacity parameterization,
the ``cross-term'' for lepton diffusion is absent, constraining
lepton evolution to depend only on gradients in neutrino 
chemical potential denoted, here by a single diffusion 
frequency $\omega_{\rm L,diff}.$     

Now, in the short-wavelength high-$k$ rapidly diffusing limit, the 
characteristic frequency associated with lepton diffusion satisfies
\be
|\omega_{\rm L,diff}|\rightarrow\, 
\frac{k^2{\mathcal K}}
{\pi^2\,\kappa_0\rho\,n\left(\partial Y_L/\partial\mu_{\nu_e}\right)_{\rho,T}
}  \gg\omega.
\ee
With this, the neutrino chemical potential is approximately 
given by
\be
\delta\mu_{\nu_e}\simeq 
i\frac{\pi^2\kappa_0\rho}
{k^2{\mathcal K}}\times\left[n\,\omega
\left(\frac{\partial Y_L}{\partial\rho}
\right)_{T,\mu_{\nu_e}}
+\frac{{\bf k}\cdot{\bf F_L}}{\rho}\right]\delta\rho.
\label{deltamunu}
\ee
The first term on the right hand side produces damping in 
much the same way as classical Silk damping (Silk, 1968).  
There are however, some subtleties to this concept -- 
a point we will consider later on.  The second term
$\propto {\bf k}\cdot{\bf F_L}$ is in part, responsible for 
driving arising from the background {\it lepton} flux 
interacting with changes in the fluid's density.
Note that we have made use of the fact that $|\delta{\bf v}|
\propto\delta\rho$ for acoustic and magnetosonic waves, which 
allowed us to drop the term in eq. (\ref{deltaetalong})
$\propto\delta{\bf v}\cdot\bnabla Y_L$.    

In order to calculate $\delta T$ in terms of $\delta\rho$, we
perform the analysis for rapid diffusion with respect to the first 
law of thermodynamics.  We have
\be
-i\omega\Biggl\{\left(\frac{\partial U}
{\partial T}\right)_{\mu_{\nu_e},\rho}
\delta T
+\left(\frac{\partial U}{\partial\mu_{\nu_e}}\right)_{T,\rho}
\delta\mu_{\nu_e}&+& \left[\left(\frac{\partial U}
{\partial\rho}\right)_{T,
\mu_{\nu_e}}-\frac{U+P}{\rho}\right]\delta\rho\Biggr\}
+\delta{\bf v}\cdot\left[\bnabla U-\left(U+P\right)
\bnabla\,{\rm ln}\rho \right]\nonumber\\
&=&i\,{\bf k}\cdot{\bf F}\frac{\delta\rho}{\rho}-
k^2\frac{{\mathcal K}
k^2_BT}{\kappa_{\rm eff}\rho}\delta T-k^2
\frac{{\mathcal K}\mu_{\nu_e}}{\pi^2\kappa_0
\rho}\,\delta\mu_{\nu_e}.
\ee
Now, we may define another characteristic neutrino diffusion 
frequency, $\omega_{\rm E,diff}$ which measures the relative
importance of energy transport via the diffusion of neutrinos
down their temperature gradient
\be
\omega_{\rm E,diff}\equiv \omega+
i\frac{k^2{\mathcal K}k^2_BT}{\kappa_{\rm eff}\rho
\,\left(\partial U/\partial T\right)_{\rho,\mu_{\nu_e}}}.
\ee
In a similar fashion to $\delta\mu_{\nu_e}$, the temperature 
perturbation becomes
\be
\delta T&=&-\frac{\omega}{\omega_{\rm E,diff}}
\frac{\left(\partial U/\partial\mu_{\nu_e}\right)_{\rho,T}}
{\left(\partial U/\partial T\right)_{\rho,\mu_{\nu_e}}}\,
\delta\mu_{\nu_e}
-\frac{\omega}{\omega_{\rm E,diff}}
\frac{\left(\partial U/\partial\rho\right)_{T,\mu_{\nu_e}}-
\left(U+P\right)/\rho}{\left(\partial U/\partial T\right)_
{\rho,\mu_{\nu_e}}}\,\delta\rho
-i\frac{\delta{\bf v}\cdot\left[\bnabla U
-\left(U+P\right)\bnabla{\rm ln}\rho\right]}{\omega_{\rm E,diff}
\left(\partial U/\partial T\right)_{\rho,\mu_{\nu_e}}}
\nonumber\\
&-&\frac{{\bf k}\cdot{\bf F}}{\omega_{\rm E,diff}
\left(\partial U/\partial T\right)_{\rho,\mu_{\nu_e}}}\frac{\delta
\rho}{\rho}
-i\frac{k^2{\mathcal K}\mu_{\nu_e}}{\pi^2\kappa_0\,\rho\,
\omega_{\rm E,diff}
\left(\partial U/\partial T\right)_{\rho,\mu_{\nu_e}}}\delta
\mu_{\nu_e}.
\ee
In the short-wavelength high-$k$ limit, energy transport by means of 
neutrino diffusion is rapid compared to the acoustic or magnetoacoustic
crossing time such that
\be
|\omega_{\rm E,diff}|\rightarrow \frac{k^2{\mathcal K}k^2_BT}
{\kappa_{\rm eff}\rho
\,\left(\partial U/\partial T\right)_{\rho,\mu_{\nu_e}}}\gg\omega.
\ee
Again, by realizing that $\delta\mu_{\nu_e}\propto\,k^{-1}\delta\rho$ and 
$|\delta{\bf v}|\propto\delta\rho$ for acoustic and magnetosonic waves in 
short-wavelength limit, the temperature perturbation may be written as
\be
\delta T\simeq i\frac{\kappa_{\rm eff}\rho}{k^2{\mathcal K}k^2_BT}
\times\Biggl\{
\omega\left[\left(\frac{\partial U}{\partial\rho}\right)_{T,\mu_{\nu_e}}
-\frac{U+P}{\rho}\right]\delta\rho
+\frac{{\bf k}\cdot{\bf F}}{\rho}\delta\rho 
+i\frac{k^2{\mathcal K}\mu_{\nu_e}}{\pi^2\kappa_0\rho}
\delta\mu_{\nu_e} \Biggr\},
\ee 
and with the help of eq. (\ref{deltamunu}), we have
\be
\delta T\simeq i\frac{\kappa_{\rm eff}\rho}{k^2{\mathcal K}k^2_BT}
\times\Biggl\{
\omega\left[\left(\frac{\partial U}{\partial\rho}\right)_{T,\mu_{\nu_e}}
-\frac{U+P}{\rho}-
\mu_{\nu_e}\,n\,
\left(\frac{\partial Y_L}{\partial\rho}\right)_{T,\mu_{\nu_e}}
\right]
+\frac{{\bf k}\cdot\left[{\bf F}-\mu_{\nu_e}\,{\bf F_L}
\right]}{\rho}
 \Biggr\}\,\delta\rho.
\label{deltaT}
\ee
The first three terms inside the braces in eq. (\ref{deltaT}) represent 
Silk damping.  Upon compression, neutrinos leak out along the temperature 
gradient of the fluid element and cool the gas.  Cooling during
compression can be thought of as a general criterion for damping.  
The term 
$\propto\left[{\bf F}-\mu_{\nu_e}{\bf F_L}\right]$ is responsible for 
unstable driving due to a combination of both the energy and lepton flux
coupling to fluctuations in density.       

In this light, the total pressure perturbation is divided into two parts;
one which yields a standard acoustic response $\propto\delta\rho$ 
while the other
component $\propto i\,k^{-1}\delta\rho$ leads to 
radiative driving and damping.  That is,
\be
\delta P=\delta P_{\rm ac}+\delta {\tilde P}
\ee  
where
\be
\delta P_{\rm ac}=\left(\frac{\partial P}{\partial\rho}\right)_{T,
\mu_{\nu_e}}\,\delta\rho\,\,\,\,\,\,\,&{\rm and}&\,\,\,\,\,\,\,
\delta {\tilde P}=\left(\frac{\partial P}{\partial T}\right)_{\rho,
\mu_{\nu_e}}\,\delta T +\left(\frac{\partial P}{\partial 
\mu_{\nu_e}}\right)_{\rho,T}\,\delta\mu_{\nu_e}.
\ee
The quantity 
\be
\left(\frac{\partial P}{\partial\rho}\right)_{T,
\mu_{\nu_e}}=\left(\frac{\partial P}{\partial\rho}\right)_{T,
\mu_{\nu_e},\mu_e}+\left(\frac{\partial\mu_e}{\partial\rho}
\right)_{T,\mu_{\nu_e}}\left(\frac{\partial P}{\partial
\mu_e}\right)_{\rho,T,\mu_{\nu_e}}\equiv c^2_i
\ee
is the square of the isothermal and ``isomeric'' sound speed 
$c_i$.\footnote{By ``isomeric'' we mean that the composition of a fluid
element is fixed.  However, this is somewhat misleading.  Only when the 
chemical potentials of all of the individual particle species are
fixed can the composition be thought of as being frozen in.  In the limit of
rapid electron-type
neutrino diffusion, the chemical potential of the neutrino is fixed over 
a wavelength to lowest ${\mathcal O}(k^{-1})$.  Therefore, the relative 
concentrations of electrons, protons, and neutrons may evolve subject to 
the constraint $\delta \mu_p+\delta\mu_e-\delta\mu_n =\delta\mu_{\nu_e}$=0.}  
When pair capture and emission occur prodigiously, the rapid exchange 
of electron-type 
neutrinos fixes the gas temperature to that of the neutrinos and the
composition of the $npe\gamma$ gas is determined by the evolution
of the electron-type neutrino's chemical potential.  Due to rapid 
neutrino diffusion, which depends on the gradients of both the neutrino 
temperature and chemical potential, perturbations in 
temperature and neutrino chemical potential are smaller by a factor 
$\propto k^{-1}$ in the short-wavelength limit.  Upon 
compression, both the temperature and relative chemical  composition of a given 
fluid element remain fixed to lowest ${\mathcal O}(k^{-1})$ while 
its density and electron chemical potential evolve.  With this in mind, 
by combing eqs. (\ref{deltaP}), (\ref{deltamunu}), and (\ref{deltaT})
the total pressure perturbation may be written as
\be
\delta P=c^2_i\,\delta\rho&+&i\left(\frac{\partial P}
{\partial T}\right)_{\rho,\mu_{\nu_e}}\frac{\kappa_{\rm eff}\rho}
{k^2{\mathcal K}k^2_BT}\times\Biggl\{
\omega\left[\left(\frac{\partial U}{\partial\rho}\right)_{T,\mu_{\nu_e}}
-\frac{U+P}{\rho}-
\mu_{\nu_e}\,n\,
\left(\frac{\partial Y_L}{\partial\rho}\right)_{T,\mu_{\nu_e}}
\right]
+\frac{{\bf k}\cdot\left[{\bf F}-\mu_{\nu_e}\,{\bf F_L}
\right]}{\rho}
 \Biggr\}\,\delta\rho\nonumber\\
&+&i\left(\frac{\partial P}{\partial \mu_{\nu_e}}\right)_{\rho,T}
\frac{\pi^2\kappa_0\rho}
{k^2{\mathcal K}}\times\Biggl\{n\,\omega
\left(\frac{\partial Y_L}{\partial\rho}
\right)_{T,\mu_{\nu_e}}
+\frac{{\bf k}\cdot{\bf F_L}}{\rho}\Biggr\}\delta\rho.
\label{deltaPbig}
\ee

\subsection{A Note on The First Law of Thermodynamics in the
Limit of Rapid Neutrino Diffusion}

Here, we resort to the first law of thermodynamics 
\be
dE=TdS-PdV+\sum_i\,\mu_i\,dN_i
\ee
in order to gain insight into the 
results accumulated thus far.  It is convenient to
write the first law in terms of $U=E/V$ and $Y_L$ rather
than $E$ and $N_i$ i.e.,
\be
dU-\frac{U+P}{\rho}d\rho=n\,T\,ds
+n\,\mu_{\nu_e}\,dY_L,
\ee
where $s$ is the entropy per baryon.  By differentiating with respect to 
density
\be
n\,T\,\left(\frac{\partial s}{\partial\rho}\right)_
{T,\mu_{\nu_e}}=\left(\frac{\partial U}{\partial\rho}\right)_
{T,\mu_{\nu_e}}
-\frac{U+P}{\rho}-
\mu_{\nu_e}\,n\,
\left(\frac{\partial Y_L}{\partial\rho}\right)_{T,\mu_{\nu_e}},
\ee
we see that the right hand side yields the thermodynamic 
motivation for Silk damping, when compared to the appropriate terms
in eqs. (\ref{deltaT}) and (\ref{deltaPbig}).  If pressure
leads density in time such that $\delta P\propto -i\,\delta\rho$, 
the mode will be damped.  Thus, the criteria for Silk damping
in PNSs is given by
\be
n\,T\,\left(\frac{\partial s}{\partial\rho}\right)_
{T,\mu_{\nu_e}}=\left(\frac{\partial U}{\partial\rho}\right)_
{T,\mu_{\nu_e}}
-\frac{U+P}{\rho}-
\mu_{\nu_e}\,n\,
\left(\frac{\partial Y_L}{\partial\rho}\right)_{T,\mu_{\nu_e}}
< 0.
\ee
If the entropy of the fluid decreases upon an increase in density,
then the fluid cools when compressed and is damped out (for 
now we are only considering the homogeneous case where the 
effects of the background ${\bf F}$ and ${\bf F_L}$ are small).  Note
however, that
\be
\left(\frac{\partial U}{\partial\rho}\right)_
{T,\mu_{\nu_e}}
-\frac{U+P}{\rho}<0\,\,\,\,\,\,\,{\rm and}\,\,\,\,\,\,\,
\left(\frac{\partial Y_L}{\partial\rho}\right)_{T,\mu_{\nu_e}}
< 0.
\ee  
Hence, upon compression it is {\it possible} for radiative diffusion 
to heat the fluid rather than to cool it, making it possible
for {\it radiative diffusion to  drive compressible perturbations unstable}.  
In the $\mu_{\nu_e}\rightarrow 0$ limit, cooling necessarily 
accompanies compression as in the case of classical Silk damping, 
where neutrinos only carry heat along with them as they diffuse down a 
temperature gradient.  To show this, we again exploit the first law
\be
\frac{dU}{dt}-\frac{U+P}{\rho}\frac{d\rho}{dt}=-\bnabla\cdot{\bf F}
=n\,T\,\frac{ds}{dt}+n\,\mu_{\nu_e}\frac{dY_L}{dt}
\ee
where ${\bf F}$ denotes the flux of {\it energy}.  When the 
chemical potential of the radiation is zero, the energy flux 
mediates the transport of heat alone.  To further
illustrate this point, we re-write the first law as
\be
n\,T\,\frac{ds}{dt}&=&-\bnabla\cdot\left[{\bf F}-\mu_{\nu_e}
{\bf F_L}\right]-{\bf F_L}\cdot\bnabla\mu_{\nu_e}
\ee    
Burrows, Mazurek, and Lattimer (1981) noted that the second term on 
the right hand side is always positive for global cooling models
of PNSs and therefore, always leads to heating as neutrinos escape.
Those authors refer to to this effect as ``Joule heating'' in analogy 
with the well known effect in electrodynamics.
\footnote{This point is made clear
by realizing that ${\bf F_L}\cdot\bnabla\mu_{\nu_e}=-|{\bf F_L}|^2/\sigma$
where $\sigma$ is the radiative conductivity associated with lepton 
diffusion and that the lepton flux ${\bf F_L}$ can be thought of 
as a current density.}  Upon perturbation, the first law becomes
\be
-i\omega\,n\,T\left[\delta s+\bxi\cdot\bnabla s\right]=
-i{\bf k}\cdot\left[\delta{\bf F}-
\mu_{\nu_e}\delta{\bf F_L}\right]+\delta\mu_{\nu_e}\bnabla\cdot{\bf F_L},
\ee
where last term on the right hand side vanishes since ${\bf F_L}$ is a constant
for our equilibrium.  This leaves us with 
\be
-i\omega\,n\,T\delta s\simeq -i{\bf k}\cdot\left[\delta{\bf F}-
\mu_{\nu_e}\delta{\bf F_L}\right]=-i{\bf k}\cdot\delta{\bf F_Q}
\label{deltas}
\ee 
where $\delta {\bf F}_Q$ can be thought of as a heat flux and is 
responsible for cooling the 
perturbations via radiative diffusion. From eqs. (\ref{Ftotal})-(\ref{Flepton}),
the heat flux ${\bf F_Q}\equiv {\bf F}-\mu_{\nu_e}{\bf F_L}$ only 
depends on gradients in temperature i.e., heat is transported
via changes in temperature alone.  Burrows \& Lattimer (1986) point out
that this is due to our parameterization of opacity, which 
depends on the square of energy rather than on energy and 
composition.  Thus, 
the temperature perturbation in the limit of 
rapid neutrino diffusion  given by eq. (\ref{deltaT}) can be written
in a more compact, yet descriptive form
\be
\delta T\simeq i\frac{\kappa_{\rm eff}\rho}{k^2{\mathcal K}k^2_BT}
\times\left[
\omega\,n\,T\left(\frac{\partial s}{\partial\rho}\right)_{T,
\mu_{\nu_e}}
+\frac{{\bf k}\cdot{\bf F}_Q}{\rho}
 \right]\,\delta\rho.
\label{deltaTs}
\ee

Compare eq. (\ref{deltamunu}) with eq. (\ref{deltaTs}).  Clearly, the entropy 
content of a fluid element is tied to its temperature perturbation while its
lepton fraction is tied to fluctuations in the neutrino chemical potential.  
This
reinforces the concept that heat flows down gradients in temperature while 
degeneracy energy is transported due to gradients in neutrino chemical 
potential.

\subsection{The Possibility of Diffusive Heating of Compressible Waves}

In the limit of rapid neutrino diffusion, if the entropy of a 
compressed fluid element increases appreciably to counter the loss of 
degeneracy pressure that accompanies lepton transport, then the fluctuation
gains energy during an oscillation.  The criterion for this to occur 
and its applicability to PNSs is sketched out in Appendix \ref{a:b}.

The entropy of a fluid element increases if the baryons
exhibit some level of degeneracy as in the case of PNS cores.  In order 
to illustrate this property, compare the ``Silk determinant'' for an
ideal non-relativistic, non-degenerate gas of baryons ($P\propto \rho T$) 
with that of a completely degenerate non-interacting and non-relativistic 
gas of baryons
\be
\frac{\partial s_{\rm ideal}}{\partial\rho}=
\frac{\partial U}{\partial\rho}-\frac{U+P}{\rho}=-\frac{P}{\rho}<0
\,\,\,\,
\,\,{\rm while}\,\,\,\,\,\,\frac{\partial s_{\rm deg}}{\partial\rho}
=0.
\ee
The fact that the entropy response of a degenerate gas vanishes
follows directly from the third law of thermodynamics.  In other words, 
a purely degenerate gas is adiabatic by {\it fiat} since it contains 
zero entropy.   
Furthermore, in a mixture of baryons consisting of protons and neutrons in 
the core of PNSs, the total equation of state may be 
super-adiabatic in the sense that the net entropy increases upon compression.  
Consider the zero-point energy density of the neutrons given 
by eq. (\ref{Pfermi}).  The estimate given in Appendix \ref{a:b} indicates
that upon compression, the entropy of the neutrons
increases in the limit of rapid neutrino diffusion.  An increase in 
density necessarily leads to a decrease in lepton fraction $Y_L$ 
due to overall charge neutrality.  The removal of a proton, which
results from deleptonization, and 
the addition of a neutron results in
a net increase in pressure or energy per unit volume upon compression since 
the chemical potential i.e.,  Fermi energy, of the neutrons is greater 
than that of the protons.

Another positive contribution to the Silk determinant 
originates from changes in electron pressure or energy density resulting
from changes in density.  Though 
the electron pressure does not explicitly depend on fluid density,
it implicitly does through the electron chemical potential and the 
condition for  chemical equilibrium.  In the limit of 
large electron degeneracy, the electron pressure $U_e=3P_e\propto\mu^4_e$.
Therefore, changes in $U_e$ result from changes in $\mu_e$.  The condition 
for chemical equilibrium, eq. (\ref{beta}), informs us that $\mu_e$ 
increases by a factor of $k_BT(Y_p/Y_n)$ upon compression.  
Eq. (\ref{silkelec}) displays this point.

The deleptonization of a fluid element contributes to the Silk determinant,
quantified by the term $\propto\mu_{\nu_e}\partial Y_L/\partial\rho$
at fixed temperature $T$ and neutrino chemical potential $\mu_{\nu_e}$.  
It is tempting to equate the phenomenon of heating due to deleptonization 
for compressible waves to Joule heating.  Such a comparison would be 
misleading at best.  Rather than an energy deposition mechanism 
$\propto |{\bf F}_L|^2$, this is a purely linear phenomenon.  The Silk 
determinant, which is $\propto\left(\partial s/\partial\rho\right)$ at 
constant $T$ and $\nu_e$, denotes only the entropy evolution of a fluid 
element.  The deleptonization term merely acts to enforce that the heat 
content of the neutrinos, rather than heat and degeneracy energy, is 
being taken into account as the entropy of a fluid element changes during 
compression.

The sum of all these actions together must overcome the loss of 
degeneracy pressure accompanied by deleptonization.  Damping due to loss of 
lepton pressure is provided by the component of the pressure perturbation 
that originates from changes in chemical potential.  This effect occurs in 
global cooling models of PNSs as well (Burrows, Mazurek, \& Lattimer 
1981).  Not only is thermal energy altered by neutrino diffusion, but 
the large lepton degeneracy energy (particularly in the core) is lost as 
well.  Loss of lepton pressure resulting from neutrino diffusion always
damps out compressible motion.

\section{Coupling of Magnetosonic Motion to the Background Neutrino Flux:
Neutrino Bubbles}\label{s:mechanics}

Below we derive the stability criteria and growth rates for the 
fast and slow neutrino bubble instability while explaining the 
nature of the driving mechanism as well.  Two distinct, yet 
equivalent, interpretations of these radiative magnetoacoustic 
instabilities will be presented.  One
comes from understanding the relationship between the radiation 
pressure and density perturbations.  If radiation pressure lags behind
density (in time), heating accompanies compression -- the necessary 
criterion for driving. In the frame of the fluid element, the radiation
pressure perturbation is composed of three distinct components.  The 
balance between the three components ultimately determines the stability
properties of the magnetoacoustic mode.  Another way of understanding
radiative 
magnetoacoustic driving is by examining the relationship between 
the fluid velocity (or displacement) and the perturbed radiation force.
The radiation force perpendicular to the wave vector is 
given by the sum of the perpendicular components of the linear energy
and lepton flux perturbation.  This drives magnetosonic motion unstable since 
the fluid oscillates in part along the magnetic field line.  That is, 
the uniform background magnetic field ${\bf B}$ may in general, lie in 
any direction with respect to the wave vector ${\bf k}$, allowing the 
motion of the fluid and the radiative driving force to coincide, leading to 
a net over-stability.

We first identify the basic properties of standard magnetosonic 
waves in the absence of stratification and radiative diffusion.  In
order to understand neutrino bubble instability from the 
relationship between pressure and density perturbations, we 
calculate the 
growth rates by use of the magnetosonic wave equation and isolate 
the relevant components of the pressure perturbation that lead to
damping and driving.  We then perform the exact same analysis on 
the linearized equation of motion where the component of the  perturbed 
radiative energy and lepton flux perpendicular to the wave vector drives 
magnetosonic modes unstable.

\subsection{Basic Magnetosonic Waves}

Linearizing the continuity, momentum, and induction equations, as well
as Gauss' Law, yields
\begin{equation}
-i\omega\delta\rho + i\rho {\bf k}\cdot\delta {\bf v} +
\delta {\bf v}\cdot{\bnabla}\rho=0,
\label{dcont}
\end{equation}
\begin{equation}
-i\omega\rho\delta{\bf v}=-i{\bf k}\,\delta P+
{\bf g}\,\delta\rho+{i\over4\pi}({\bf k}\times\delta{\bf B})\times{\bf B},
\label{dgasmom}
\end{equation}
\begin{equation}
-i\omega\delta{\bf B}=i{\bf k}\times(\delta{\bf v}\times{\bf B}),
\label{dfluxfreezing}
\end{equation}
and
\be
{\bf k}\cdot\delta{\bf B}=0,
\label{dgauss}
\ee
respectively.

The procedure for obtaining the asymptotic growth rate for 
radiatively driven magnetoacoustic instabilities was carried out in 
detail by BS03.  Based on analysis of the general (and cumbersome) 
dispersion relation, they concluded that in the short wave-length limit,
the radiatively driven MHD instabilities behaved as the standard fast
and slow magnetoacoustic modes to lowest order where the effects of 
radiation produced a relatively small imaginary correction to the 
mode frequency.  In order to isolate the mechanism for over-stability, 
BS03 then performed an asymptotic modal analysis where they solved 
the momentum equation by using the standard fast and slow eigenvectors
and applied them to all of the ${\mathcal O}(k^{-1})$ terms in the 
momentum equation; the pressure driving terms $\propto\delta T$, 
buoyancy force $\propto{\bf g}\,\delta\rho$, and to the Eulerian component 
of the 
the density perturbation which is $\propto\delta{\bf v}\cdot
\bnabla{\rm ln}\rho$.  This 
technique is mathematically similar to the work integral approach 
in classic stellar
pulsation theory (Unno 1989) for obtaining growth rates due to 
non-adiabatic driving of nearly adiabatic pulsations.  

To lowest order, the linear momentum equation is given by 
\be
-i\omega\rho\delta{\bf v}=-i{\bf k}\left[\delta P_{\rm ac}+
\frac{{\bf B}\cdot\delta{\bf B}}{4\pi}\right]+
i\frac{{\bf k}\cdot{\bf B}}{4\pi}\delta{\bf B}
\label{tempeuler}
\ee   
where $\delta P_{\rm ac}\equiv c^2_{\rm i}\delta\rho$ is the short-wavelength
acoustic pressure response to a density perturbation $\delta\rho$.   Also
to lowest order,
\be
\delta\rho=\rho\frac{{\bf k}\cdot\delta{\bf v}}{\omega}\,\,\,\,\,\,\,\,
{\rm and}\,\,\,\,\,\,\,\,
\delta{\bf B}=\frac{{\bf B}}{\omega}\left({\bf k}\cdot\delta{\bf v}
\right) - \frac{{\bf k}\cdot{\bf B}}{\omega}\delta{\bf v}.
\ee
Eliminating $\delta{\bf B}$, we obtain
\be
-i\rho\frac{{\tilde\omega}^2}{\omega}\delta{\bf v}=
-i{\bf k}\left[\left(c^2_i+v^2_A\right)\delta\rho
-c^2_i\frac{\left({\bf k}\cdot{\bf v_A}\right)^2}
{\omega^2}\delta\rho\right]+i\left({\bf k}\cdot{\bf v_A}\right)
{\bf v_A}\,\delta\rho.
\label{tempeuler2}
\ee
where ${\bf v_A}={\bf B}/\sqrt{4\pi\rho}$ is the Alfv\'en velocity
and ${\tilde\omega}^2\equiv\omega^2-({\bf k}\cdot{\bf v_A})^2$.  
In the optically thick limit, neutrinos force the gas via a gradient
in their pressure which is $\parallel$ to ${\bf k}$ in the short wave-length 
limit.  This requires the waves to possess some longitudinal, or compressible, 
component which automatically rules out the purely 
incompressible shear Alfv\'en wave as a candidate for over-stable 
radiative driving.  In order to isolate the
magnetosonic fast and slow waves, let's define a mode polarization  
${\hat{\bf\epsilon}}(\omega,k)$ such that $\delta{\bf v}=
{\hat{\mathbb\epsilon}}(\omega,k)\psi(\omega,k)$ where $\psi$ is some
complex-valued amplitude.  Choosing $\psi=\delta\rho/\rho$ gives us the 
magnetosonic eigenvectors,
\be
{\hat{\bf\epsilon}}(\omega,k)=\frac{\omega}{{\tilde\omega}^2}\left[
\left(\frac{{\tilde\omega}^2}{\omega^2}c^2_i+v^2_A\right){\bf k}-
\left({\bf k}\cdot{\bf v_A}\right){\bf v_A}\right]
\label{polarization}
\ee 
where the two magnetosonic modes must satisfy the dispersion 
relation
\be
\omega^2=k^2\,c^2_i\frac{{\tilde\omega}^2}{\omega^2}+k^2v^2_A.
\label{dispersion}
\ee
In principle, the fast and slow modes posses components both 
$\parallel$ and $\perp$ to 
${\bf k}$.  This mixed longitudinal and transverse behavior
is central to the MHD radiative driving mechanism (BS03), a point we will
now discuss.

\subsection{The Magnetosonic Wave Equation Subject to
Rapid Radiative Diffusion:  Asymptotic Growth Rate}

Taking the divergence of eq. (\ref{dgasmom}), immediately allows us to
restrict our analysis to compressible short-wavelength motions i.e.,
the modes that can be driven unstable by the effects of radiative 
diffusion.  We have
\be
-i\omega{\bf k}\cdot\delta{\bf v}=-ik^2\left[\frac{\delta P}{\rho}
+\frac{{\bf B}\cdot\delta{\bf B}}{4\pi\rho}\right] +\left({\bf k}\cdot
{\bf g}\right)\frac{\delta\rho}{\rho}
\ee
where we have made use of eq. (\ref{dgauss}).  Eliminating $\delta{\bf B}$  
using eq. (\ref{dfluxfreezing}) gives us
\be
-i\omega\left({\bf k}\cdot\delta{\bf v}\right)=-i\,k^2\left[
\frac{\delta P}{\rho}+v^2_A\frac{\left({\bf k}\cdot\delta
{\bf v}\right)}{\omega} -\frac{\left({\bf k}\cdot
{\bf v_A}\right)^2}{\omega^2}\frac{\delta P}{\rho}
-i\frac{\left({\bf g}\cdot{\bf v_A}\right)\left({\bf k}
\cdot{\bf v_A}\right)}{\omega^2}\frac{\delta\rho}{\rho} \right]
+\left({\bf k}\cdot{\bf g}\right)\frac{\delta\rho}{\rho}.
\ee 
Dividing the pressure perturbation into its acoustic $\delta P_{\rm ac}$ and 
driving/damping $\tilde{\delta P}$ contributions and by
utilizing eq. (\ref{dcont}), the magnetosonic wave equation becomes
\be
\left[\omega^2-k^2\frac{{\tilde\omega}^2}{\omega^2}c^2_i
-k^2v^2_A\right]\frac{\delta\rho}{\rho}\simeq -i\omega\delta
{\bf v}\cdot\bnabla{\rm ln}\rho+k^2\Biggl\{\frac{{\tilde\omega}^2}
{\omega^2}\frac{\delta{\tilde P}}{\rho}+iv^2_A\frac{\delta
{\bf v}}{\omega}\cdot\bnabla{\rm ln}\rho -i\frac{\left({\bf g}
\cdot{\bf v_A}\right)\left({\bf k}
\cdot{\bf v_A}\right)}{\omega^2}\frac{\delta\rho}{\rho} \Biggr\}
+i\left({\bf k}\cdot{\bf g}\right)\frac{\delta\rho}{\rho}.
\label{sonic1}
\ee
Neglecting terms depending on background gradients or those originating
from radiative diffusion leaves us with only the left hand side of eq. 
(\ref{sonic1}) i.e., the magnetosonic wave equation for a 
uniform background.  The terms on the right hand side 
of eq. (\ref{sonic1}) produce ${\mathcal O}(k^{-1})$ corrections to a 
magnetosonic wave and are responsible for the  growth and damping 
rates which can be viewed as ${\mathcal O}(k^{-1})$ corrections 
to the mode frequency.  In order to evaluate the frequency correction 
arising from radiative driving, we let
$\omega\rightarrow\omega +A_0$, where $A_0$ is the damping or driving rate
which is independent of the wavenumber $k$.

It is useful to recast the gravitational acceleration, ${\bf g}$, 
by using hydrostatic balance 
\be
{\bf g}=\frac{1}{\rho}\bnabla P=c^2_i\bnabla{\rm ln}\rho+\frac{1}{\rho}
\left[\left(\frac
{\partial P}{\partial T}\right)_{\rho,\mu_{\nu_e}}
\bnabla T + \left(\frac{\partial P}{\partial\mu_{\nu_e}}\right)_{\rho,T}
\bnabla\mu_{\nu_e}\right].
\label{hydrostatic}
\ee 
This, along with the magnetosonic polarization vector given 
by eq. (\ref{polarization}) allows us to eliminate all of the 
terms $\propto\bnabla{\rm ln}\rho$ in eq. (\ref{sonic1}).  The asymptotic 
prescription for the eigenvalues mentioned 
above allows us to convert eq. (\ref{sonic1}) into an expression for the 
damping or driving rate.  After a bit of algebra, we have
\be
\frac{2\,A_0}{\omega}\left[2\omega^2-k^2\left(c^2_i+
v^2_A\right)\right]\frac{\delta\rho}{\rho}&\simeq &k^2\Biggl\{ 
\frac{{\tilde\omega}^2}{\omega^2}\frac{\delta{\tilde P}}{\rho}
-i\frac{\left({\bf k}\cdot{\bf v_A}\right)}{\omega^2\rho}\,{\bf v_A}
\cdot\left[\left(\frac
{\partial P}{\partial T}\right)_{\rho,\mu_{\nu_e}}
\bnabla T + \left(\frac{\partial P}{\partial\mu_{\nu_e}}
\right)_{\rho,T}\bnabla\mu_{\nu_e}\right]\frac{\delta\rho}{\rho}
\Biggr\}\nonumber\\
&+&i\frac{{\bf k}}{\rho}\cdot\left[ \left(\frac
{\partial P}{\partial T}\right)_{\rho,\mu_{\nu_e}}
\bnabla T + \left(\frac{\partial P}{\partial\mu_{\nu_e}}
\right)_{\rho,T}\bnabla\mu_{\nu_e}\right]\frac{\delta\rho}{\rho}.
\label{sonic2}
\ee  
If we expand $\delta{\tilde P}$ by use of eq. (\ref{deltaPbig}) while
noting that 
\be
\bnabla T=-\frac{\kappa_{\rm eff}\,\rho}{{\mathcal K}k^2_BT}\left[
{\bf F}-\mu_{\nu_e}{\bf F_L}\right]\,\,\,\,\,\,\,\,\,\,\, {\rm and}
\,\,\,\,\,\,\,\,\,\,\,\bnabla\mu_{\nu_e}=-\frac{\pi^2\kappa_0\,\rho}{
{\mathcal K}}{\bf F_L},
\label{gradients}
\ee
the growth rate $A_0$ from eq. (\ref{sonic2}) is given by
\be
A_0&=&\frac{i}{2\left[2\omega^2-
k^2\left(c^2_i+v^2_A\right)\right]}\times\Biggl\{
\frac{\pi^2\kappa_0}{{\mathcal K}}\left(\frac{\partial P}
{\partial\mu_{\nu_e}}\right)_{\rho,T}\left[\rho{\tilde\omega}^2
n\left(\frac{\partial Y_L}{\partial\rho}\right)_{\rho,\mu_{\nu_e}}
+\frac{\left({\bf k}\cdot{\bf v_A}\right)}{\omega}\left({\bf k}
\times{\bf v_A}\right)\cdot\left({\bf k}\times{\bf F_L}\right)\right]
\nonumber\\
&+&\frac{\kappa_{\rm eff}}{{\mathcal K}k^2_BT}\left(
\frac{\partial P}{\partial T}\right)_{\rho,\mu_{\nu_e}}\left[\rho
{\tilde{\omega}}^2\left(\left(\frac{\partial U}{\partial\rho}\right)_{T,
\mu_{\nu_e}}-\frac{U+P}{\rho}-\mu_{\nu_e}n\left(\frac{\partial Y_L}
{\partial\rho}\right)_{T,\mu_{\nu_e}}\right)+\frac{\left({\bf k}\cdot
{\bf v_A}\right)}{\omega}\left({\bf k}\times{\bf v_A}\right)\cdot
\left({\bf k}\times\left[{\bf F}-\mu_{\nu_e}{\bf F_L}\right]\right)
\right]
\Biggr\}.
\label{growthrate}
\ee

\subsection{Analogy with the Work Integral}

BS03 noticed that when the components of the radiation pressure 
perturbation originating from gradients in background 
quantities are non-zero, radiative driving may then occur.  Consider the 
Lagrangian perturbation of the total neutrino pressure defined as 
\be
\Delta P_{\nu}\equiv\Delta\left[P_{\nu_e}+P_{\bar{\nu_e}}+
P_{\nu_X}+P_{{\bar{\nu_X}}}\right]={\bxi}\cdot\bnabla P_{\nu}+
\delta P_{\nu}
\ee
where $\bxi=i\delta{\bf v}/\omega$ is the fluid's Lagrangian 
displacement.  More explicitly, 
\be
\Delta P_{\nu}=\left(\frac{\partial P_{\nu}}
{\partial T}\right)_{\mu_{\nu_e}}\bxi\cdot
\bnabla T +
\left(\frac{\partial P_{\nu}}{\partial\mu_{\nu_e}}
\right)_{T}\bxi\cdot\bnabla\mu_{\nu_e}+\left(
\frac{\partial P_{\nu}}{\partial T}\right)_{\mu_{\nu_e}}
\delta T+\left(\frac{\partial P_{\nu}}{\partial\mu_{\nu_e}}
\right)_{T}\delta\mu_{\nu_e}.
\ee
The density dependence in the pressure 
derivatives is discarded since $P_{\nu}$ depends only on 
temperature and neutrino chemical potential.  By using eqs. 
(\ref{deltamunu}), (\ref{deltaT}), and (\ref{gradients}), 
the driving component of $\Delta P_{\nu}$ becomes
\be
\Delta P^{\rm drive}_{\nu}
&=&-\left(\frac{\partial P_{\nu}}
{\partial T}\right)_{\mu_{\nu_e}}\frac{\kappa_{\rm eff}
\rho}{{\mathcal K}k^2_BT\,k^2}\Biggl\{k^2\,\bxi\cdot\left[{\bf F}-
\mu_{\nu_e}{\bf F_L}\right]-
{\bf k}\cdot\left[{\bf F}-\mu_{\nu_e}{\bf F_L}
\right]\left({\bf k}\cdot\bxi\right)\Biggr\}
-\left(\frac{\partial P_{\nu}}
{\partial\mu_{\nu_e}}\right)_{T}\frac{\pi^2\kappa_0
\rho}{{\mathcal K}\,k^2}\Biggl\{k^2\,\bxi\cdot{\bf F_L}-
{\bf k}\cdot{\bf F_L}\left({\bf k}\cdot\bxi\right)\Biggr\}\nonumber\\
&=&-\left(\frac{\partial P_{\nu}}
{\partial T}\right)_{\mu_{\nu_e}}\frac{\kappa_{\rm eff}\,
\rho}{{\mathcal K}k^2_BT\,k^2}\Biggl\{\left({\bf k}\times\bxi
\right)\cdot\left({\bf k}\times\left[{\bf F}-\mu_{\nu_e}{\bf F_L}
\right]\right)\Biggr\}
-\left(\frac{\partial P_{\nu}}
{\partial\mu_{\nu_e}}\right)_{T}\frac{\pi^2\kappa_0\,
\rho}{{\mathcal K}\,k^2}\Biggl\{\left({\bf k}\times\bxi\right)
\cdot\left({\bf k}\times{\bf F_L}\right)\Biggr\},
\label{DeltaP1}
\ee   
and by using the magnetosonic eigenvector given by eq. 
(\ref{polarization}), the driving component becomes
\be
\Delta P^{\rm drive}_{\nu}&=&i\frac{\left({\bf k}\cdot
{\bf v_A}\right)}{{\tilde\omega}^2}
\left(\frac{\partial P_{\nu}}
{\partial T}\right)_{\mu_{\nu_e}}\frac{\kappa_{\rm eff}\,
\rho}{{\mathcal K}k^2_BT\,k^2}\Biggl\{\left({\bf k}\times{\bf v_A}
\right)\cdot\left({\bf k}\times\left[{\bf F}-\mu_{\nu_e}{\bf F_L}
\right]\right)\Biggr\}\frac{\delta\rho}{\rho}\nonumber\\
&+&i\frac{\left({\bf k}\cdot
{\bf v_A}\right)}{{\tilde\omega}^2} \left(\frac{\partial P_{\nu}}
{\partial\mu_{\nu_e}}\right)_{T}\frac{\pi^2\kappa_0\,
\rho}{{\mathcal K}\,k^2}\Biggl\{\left({\bf k}\times{\bf v_A}
\right)
\cdot\left({\bf k}\times{\bf F_L}\right)\Biggr\}
\frac{\delta\rho}{\rho}.
\label{DeltaP2}
\ee
If the radiation pressure maximum is attained after the fluid's density 
maximum, there will be a net amount of work during 
the compression cycle.  Thus, the criterion for instability via 
radiative driving can be written as 
\be
{\rm Im}\left[\Delta\rho^*\Delta P_{\nu}\right]>0
\ee
where $\Delta\rho=\delta\rho+\bxi\cdot\bnabla\rho$ is the
Lagrangian density perturbation.  It can be checked that upon 
substitution of eq. (\ref{DeltaP2}) 
into the above expression, the same criteria over-stability
is obtained as that given by eq. (\ref{growthrate}).

To gain a deeper understanding of the driving mechanism, consider 
purely hydrodynamic acoustic motion where ${\bf B}={\bf v_A}=0$ 
such that $\bxi\parallel{\bf k}$.  Substituting an appropriate 
sound wave eigenvector into eq. (\ref{DeltaP1}), we see that
\be
\Delta P^{\rm drive}_{\nu}\propto\left(\frac{\partial P_{\nu}}
{\partial T}\right)_{\mu_{\nu_e}}\frac{\kappa_{\rm eff}\,
\rho}{{\mathcal K}k^2_BT\,k^2}\Biggl\{k^2\,{\bf k}
\cdot\left[{\bf F}-
\mu_{\nu_e}{\bf F_L}\right]-
{\bf k}\cdot\left[{\bf F}-\mu_{\nu_e}{\bf F_L}
\right]\left({\bf k}\cdot{\bf k}\right)\Biggr\}
+\left(\frac{\partial P_{\nu}}
{\partial\mu_{\nu_e}}\right)_{T}\frac{\pi^2\kappa_0\,
\rho}{{\mathcal K}\,k^2}\Biggl\{k^2\,{\bf k}\cdot{\bf F_L}-
{\bf k}\cdot{\bf F_L}\left({\bf k}\cdot{\bf k}\right)\Biggr\}
=0.
\ee
That is, for purely hydrodynamic sound waves, radiative driving 
vanishes due to an exact cancellation of two unique effects, which we 
describe below.

Regardless of the direction in which a fluid oscillates, the background radiative
fluxes ${\bf F}$ and ${\bf F}_L$ always force a fluid element if 
there is a change in density.  We call this effect ``shadowing,'' 
following the terminology of the O star line-driven wind literature (Lucy
\& White 1980, Lucy 1982).  For the sake of exposition, let's ignore lepton 
number and assume that ${\bf F_L}=\mu_{\nu_e}=0$.  The component of the 
Lagrangian pressure perturbation due to shadowing originates from 
the $\bnabla\cdot\delta{\bf F}$ term in the first law of thermodynamics 
and is present in eq. (\ref{deltaPbig}) i.e.,
\be
\Delta P^{\rm shadow}_{\nu}=i\left(\frac{\partial P_{\nu}}
{\partial
T}\right)\frac{\kappa_{\rm eff}}{k^2{\mathcal K}k^2_BT}\left({\bf k}
\cdot{\bf F}\right)\delta\rho= -i\left(\frac{\partial P_{\nu}}
{\partial T}\right)\frac{{\bf k}\cdot\bnabla T}{k^2}\frac{\delta\rho}
{\rho}=-\left(\frac{\partial P_{\nu}}
{\partial T}\right)\frac{{\bf k}\cdot\bnabla T}{k^2}\left({\bf k}\cdot
\bxi\right) .
\label{shadow}
\ee
Radiation piles up 
before the density maximum 
as the local decrease in mean free path temporarily hinders the transmission of 
quanta (Baker \& Kippenhahn 1962).  However, the total extinction over a 
wavelength remains the same since a fluid element's mass is a constant 
during the compression cycle.
In order to conserve  quanta, a deficit of radiation, or a shadow, appears
upstream of the density maximum. The surplus
of radiation downstream  and the deficit of radiation upstream of the 
density maximum induces a pressure differential that peaks at the point of 
maximum density along the wave.

The contribution to the Lagrangian radiation pressure 
perturbation due to changes in the displacement vector $\bxi$ is given by
\be
\Delta P^{\rm displace}_{\nu}=\bxi\cdot\bnabla P_{\nu}=\left(\frac{\partial 
P_{\nu}}{\partial T}\right)\bxi\cdot\bnabla T.  
\label{displace}
\ee
When a fluid element moves upward along the direction of ${\bf F}$, 
the background pressure it samples is smaller than the value at its
original position.  Conversely, when a fluid element moves downward, the 
radiation pressure perturbation increases.  As previously mentioned, 
if $\bxi\parallel{\bf k}$
as in the case of a purely hydrodynamic sound waves, the pressure
perturbations from shadowing and motion along the temperature gradient
cancel one another out, a point clearly seen when comparing eqs. 
(\ref{shadow}) and (\ref{displace}).  

This harmony between $\Delta P^{\rm shadow
}_{\nu}$ and $\Delta P^{\rm displace}_{\nu}$ is broken when the 
magnetic field is introduced. The change in radiation pressure 
due to a change in position does not in general match the acoustic
response followed by a change in density since the fluid element oscillates
in part along the direction of the equilibrium ${\bf B}$. 
 
\subsection{Limit of Blaes \& Socrates 2003}

The instability mechanism considered in BS03 is 
essentially the same as in this work.  The 
major difference was their assumption of zero chemical potential for the 
photon radiation field as they were considering backgrounds that were close
to perfect black bodies.  Setting $\mu_{\nu_e}=\left(\partial P/
\partial\mu_{\nu_e}\right)=0$ in eq. (\ref{growthrate}) we have
\be
A_0=\frac{i}{2\left[2\omega^2-
k^2\left(c^2_i+v^2_A\right)\right]}\times\Biggl\{
\frac{\kappa_{\rm eff}}{{\mathcal K}k^2_BT}\left(
\frac{\partial P}{\partial T}\right)_{\rho,\mu_{\nu_e}}\left[\rho
{\tilde{\omega}}^2\left(\left(\frac{\partial U}{\partial\rho}\right)_{T,
\mu_{\nu_e}}-\frac{U+P}{\rho}\right)+\frac{\left({\bf k}\cdot
{\bf v_A}\right)}{\omega}\left({\bf k}\times{\bf v_A}\right)\cdot
\left({\bf k}\times{\bf F}\right)\right]\Biggr\},
\ee
which is completely equivalent to the one-temperature MHD growth rates
given by eqs. (93) and (107) of BS03. 

\subsection{Magnetosonic Waves Forced by Radiative Diffusion:
The Perpendicular Radiation Force}

An alternative, and more visual, physical picture for these instabilities
arises from considering forces on oscillating fluid elements.  For this 
particular application, it is both helpful and revealing 
to adopt the Lagrangian 
displacement $\bxi$ as our fundamental fluid variable.  
Other important fluid quantities are related to 
$\bxi$ in the following way
\be
\delta{\bf B}=\bnabla\times\left(\bxi\times{\bf B}\right)=
i{\bf k}\times\left(\bxi\times{\bf B}\right),\,\,\,\,\,\,\,\,\,\,
\frac{\delta\rho}{\rho}=-i\left({\bf k}
\cdot\bxi\right)-\bxi\cdot\bnabla{\rm ln}\rho,\,\,\,\,\,\,\,\,\,\,
{\rm and}\,\,\,\,\,\,\,\,\,\,\frac{\delta P_{\rm ac}}{\rho}=
-i\,c^2_i\left({\bf k}\cdot\bxi\right)-c^2_i\bxi\cdot\bnabla{\rm ln}
\rho.
\ee
This allows us to write the Euler equation, eq. (\ref{dgasmom}) in terms
of Lagrangian displacements
\be
\frac{\partial^2\bxi}{\partial t^2}\simeq-\left({\bf k}\cdot{\bf v_A}
\right)^2\bxi-{\bf k}\left(c^2_i+v^2_A\right)\left({\bf k}\cdot
\bxi\right)+{\bf v_A}\left({\bf k}\cdot{\bf v_A}\right)\left({\bf k}
\cdot\bxi\right)+{\bf k}\left({\bf k}\cdot{\bf v_A}\right)
\left(\bxi\cdot{\bf v_A}\right)
-i{\bf k}\left[\frac{\delta{\tilde P}_{\rm damp}}
{\rho}+\frac{\delta {\tilde P}_{\rm drive}}{\rho}
-c^2_i\bxi\cdot\bnabla{\rm ln}\rho\right]-i\,{\bf g}
\left({\bf k}\cdot\bxi\right)
\label{LagrangeEOM}
\ee
where $\delta{\tilde P}=\delta{\tilde P}_{\rm drive}+\delta
{\tilde P}_{\rm damp}$ and from eq. (\ref{deltaPbig})
\be
\delta{\tilde P}_{\rm drive}\equiv i\left(\frac{\partial P}
{\partial T}\right)_{\rho,\mu_{\nu_e}}\frac{\kappa_{\rm eff}\rho}
{k^2{\mathcal K}k^2_BT}\left(
{\bf k}\cdot\left[{\bf F}-\mu_{\nu_e}\,{\bf F_L}
\right]\right)
\,\frac{\delta\rho}{\rho}
+i\left(\frac{\partial P}{\partial \mu_{\nu_e}}\right)_{\rho,T}
\frac{\pi^2\kappa_0\rho}  
{k^2{\mathcal K}}\left({\bf k}\cdot{\bf F_L}\right)
\frac{\delta\rho}{\rho}.
\ee
The lowest order acoustic and magnetic tension 
restoring force terms on the right hand side of eq. (\ref{LagrangeEOM})
can be reduced to 
\be
\left({\bf k}\cdot{\bf v_A}
\right)^2\bxi+{\bf k}\left(c^2_i+v^2_A\right)\left({\bf k}\cdot
\bxi\right)-{\bf v_A}\left({\bf k}\cdot{\bf v_A}\right)\left({\bf k}
\cdot\bxi\right)-{\bf k}\left({\bf k}\cdot{\bf v_A}\right)
\left(\bxi\cdot{\bf v_A}\right)=k^2v^2_{\rm ph}\bxi
\ee
by use of eqs. (\ref{polarization})
and (\ref{dispersion}) where $v^2_{\rm ph}\equiv\omega^2/k^2$ is the 
square of the phase velocity.
The local gravitational acceleration ${\bf g}$ can be expressed in terms of
the radiative and lepton fluxes by use of hydrostatic balance 
given by eq. (\ref{hydrostatic})  and eq. (\ref{gradients})
such that
\be
{\bf g}=c^2_i\bnabla{\rm ln}\rho-\left(\frac{\partial P}{\partial T}
\right)_{\rho,\mu_{\nu_e}}\frac{\kappa_{\rm eff}}{{\mathcal K}
k^2_BT}\left[{\bf F}-\mu_{\nu_e}{\bf F_L}\right]-
\left(\frac{\partial P}{\partial\mu_{\nu_e}}
\right)_{\rho,T}\frac{\pi^2\kappa_0}{{\mathcal K}}{\bf F_L}.
\ee
Inserting this into the equation of motion yields 
\be
\frac{\partial^2\bxi}{\partial t^2}+k^2v^2_{\rm ph}\bxi&\simeq&
-i{\bf k}\frac{\delta{\tilde P}_{\rm damp}}{\rho}-i\,{\bf k}
\frac{\left({\bf k}\cdot\bxi\right)}{k^2}{\bf k}\cdot
\left[\left(\frac{\partial P}{\partial T}
\right)_{\rho,\mu_{\nu_e}} \frac{\kappa_{\rm eff}}
{{\mathcal K}k^2_BT}\left[{\bf F}-\mu_{\nu_e}{\bf F_L}\right] 
+ \left(\frac{\partial P}{\partial\mu_{\nu_e}}
\right)_{\rho,T}\frac{\pi^2\kappa_0}{{\mathcal K}}{\bf F_L}\right]
\nonumber\\
&+&i\left({\bf k}\cdot\bxi\right)
\left[\left(\frac{\partial P}{\partial T}
\right)_{\rho,\mu_{\nu_e}} \frac{\kappa_{\rm eff}}
{{\mathcal K}k^2_BT}\left[{\bf F}-\mu_{\nu_e}{\bf F_L}\right] 
+ \left(\frac{\partial P}{\partial\mu_{\nu_e}}
\right)_{\rho,T}\frac{\pi^2\kappa_0}{{\mathcal K}}{\bf F_L}\right]
+i\,{\bf k}c^2_i
\bxi\cdot\bnabla{\rm ln}\rho-i\,\left({\bf k}\cdot\bxi\right)c^2_i
\bnabla{\rm ln}\rho
\label{forcetemp}.
\ee

\begin{figure}[b!]
\begin{center}
\input{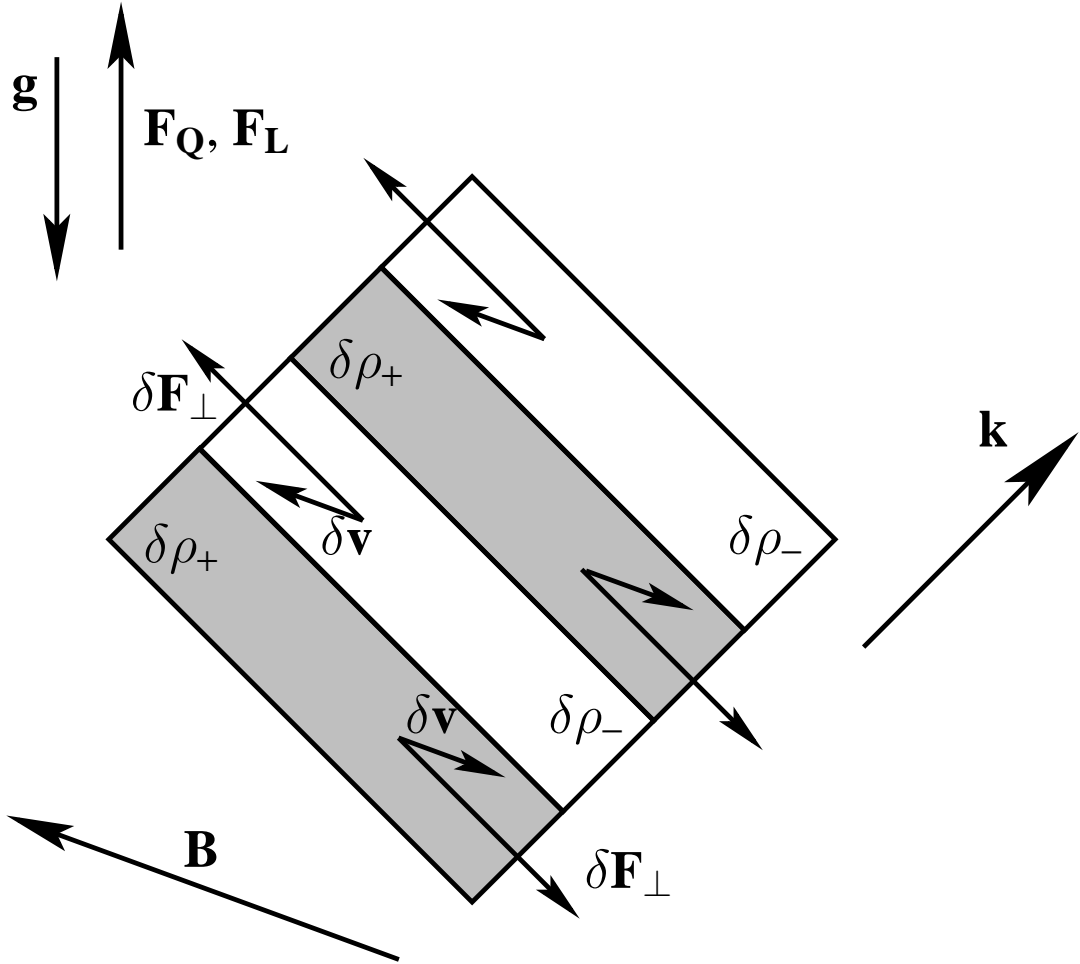}
\caption{Geometry of the neutrino bubble instability.  Here, $\delta
{\bf F_{\perp}}$ is any general perpendicular radiative flux, which 
denotes the direction of both $\delta{\bf F_L}$ and $\delta{\bf F_Q}$ 
-- assuming that the equilibrium ${\bf F_Q}$ and ${\bf F_L}$ lie in the 
same direction.  For the sake of clarity, the polarization properties of the 
depicted wave are that of the slow magnetosonic wave in the limit 
that $v^2_A\gg c^2_i$.  As a result of this choice, the velocity 
perturbation is almost perfectly $\parallel$ to the equilibrium 
magnetic field.  The driving force is $\parallel\delta{\bf F_{\perp}}$ and
possesses a finite positive projection with the velocity 
perturbation $\delta {\bf v}$, allowing for the stratified background 
radiation field to perform work on the oscillating fluid.}
\end{center}
\label{f:Omer's_wave}
\end{figure}

The driving terms on the right hand side of eq. (\ref{forcetemp}) 
may be recast in the following form
\be
-i\left(\frac{\partial P}{\partial T}\right)_{\rho,\mu_{\nu_e}}
\frac{\kappa_{\rm eff}}{{\mathcal K}k^2_BT}\left(\frac{{\bf k}\,
{\bf k}}
{k^2}-{\bf 1} \right)\cdot\left[{\bf F}-\mu_{\nu_e}{\bf F_L}
\right]\left({\bf k}\cdot\bxi\right)-i\left(\frac{\partial P}
{\partial\mu_{\nu_e}}
\right)_{\rho,T}\frac{\pi^2\kappa_0}{{\mathcal K}}\left(\frac{{\bf k}
\,{\bf k}}
{k^2}-{\bf 1} \right)\cdot{\bf F_L}\left({\bf k}\cdot\bxi\right)
+i\,c^2_{\rm i}\,\bxi\times\left({\bf k}\times\bnabla{\rm ln}\rho\right)
\label{driving}
\ee   
where the last term is orthogonal 
to to fluid displacements and therefore cannot drive the mode.
Ultimately, the linear radiation force per unit mass 
$\delta {\mathcal F}_{\rm rad}$ 
is given by a gradient in the total neutrino pressure 
$=-\frac{1}{\rho}\bnabla\left[P_{\nu_e}+P_{\bar{\nu_e}}+P_{\nu_X}\right]$ 
and upon perturbation
\be
\delta{\mathcal F}_{\rm rad}=\frac{1}{\rho}\bnabla P_{\nu}
\frac{\delta\rho}{\rho}-i\frac{{\bf k}}{\rho}\delta P_{\nu}.
\ee
where $P_{\nu}=P_{\nu_e}+P_{\bar{\nu_e}}+P_{\nu_X}$.
Following the discussion surrounding eq. (\ref{deltas}), the first term
may be broken into a component
which carries heat and one that carries lepton number
\be
\frac{1}{\rho}\bnabla P_{\nu}
\frac{\delta\rho}{\rho}=
i\left(\frac{\partial P_{\nu}}{\partial T}\right)_{\rho,
\mu_{\nu_e}}\frac{\kappa_{\rm eff}}{{\mathcal K}k^2_BT}
{\bf F_Q}\left({\bf k}\cdot
\bxi\right)+i\left(\frac{\partial P_{\nu}}{\partial\mu_{\nu_e}}
\right)_{\rho,T}\frac{\pi^2\kappa_0}{{\mathcal K}}{\bf F_L}
\left({\bf k}\cdot\bxi\right)
\ee
where we have kept the leading terms $\delta\rho$, the linear 
driving force $\delta {\mathcal F}_{\rm drive}$ can be written in 
terms of the two perturbed fluxes
\be
\delta{\mathcal F}_{\rm drive}=
{\hat{\bf k}}\times\delta{\mathcal F}_{\rm rad}=
\hat{\bf k}\times
\left[\left(\frac{\partial P_{\nu}}{\partial T}\right)_{\rho,
\mu_{\nu_e}}\frac{\kappa_{\rm eff}}{{\mathcal K}k^2_BT}\,\delta
{\bf F_Q}+\left(\frac{\partial P_{\nu}}{\partial\mu_{\nu_e}}
\right)_{\rho,T} \frac{\pi^2\kappa_0}{{\mathcal K}}
\, \delta{\bf F_L} \right].
\ee
Thus, the component of the perturbed heat and lepton 
flux $\perp$ to ${\bf k}$ leads 
to secular driving of the magnetoacoustic waves (see Figure 
2).   Due to magnetic 
tension, the fast and slow magnetoacoustic waves possess polarizations
which have components along ${\bf B}$, which in general, are {\it not}
$\parallel$ to ${\bf k}$.  Note that $\delta{\mathcal F}_{\rm drive}$ 
is the driving 
force originating from radiation pressure alone.  As previously 
argued, changes in pressure due to temperature and neutrino chemical 
potential perturbations contribute to driving.  Below the neutrinosphere,
weak interactions occur so rapidly that the temperature and composition of 
the gas are dictated by the diffusive behavior of the neutrinos.  With
the help of eq. (\ref{driving}), the total driving force becomes
\be
\delta{\mathcal F}_{\rm total}=\frac{\hat{\bf k}}{c}\times
\left[\left(\frac{\partial P}{\partial T}\right)_{\rho,
\mu_{\nu_e}}\frac{\kappa_{\rm eff}}{{\mathcal K}k^2_BT}\,\delta
{\bf F_Q}+\left(\frac{\partial P}{\partial\mu_{\nu_e}}
\right)_{\rho,T} \frac{\pi^2\kappa_0}{{\mathcal K}}
\, \delta{\bf F_L} \right].
\ee 
Both $\delta{\bf F_Q}$ and $\delta{\bf F_L}$ are $\propto -\delta
\rho/\rho$ implying that the perpendicular radiative fluxes reach their
maximum possible upward values during density minima.  Upon expansion radiation 
from the equilibrium heat and lepton fluxes more readily leaks out
perpendicular to the direction of wavefronts as to avoid regions
of enhanced density.  Due to magnetic tension, 
then fluid oscillates in part along ${\bf B}$, which in general has some
projection $\perp$ to the wavefronts.  Thus, the  
radiative driving force and the fluid velocity may oscillate in 
phase for certain propagation directions, ultimately allowing 
driving to occur.

\subsection{Estimates of Local Stability and Driving}
\label{sec:stability}

We have shown that in the limit of rapid neutrino diffusion,
magnetoacoustic waves can be radiatively driven by both a background 
${\bf F_Q}$ and ${\bf F_L}$.  Likewise, the waves can be damped by 
radiative diffusion due to the conversion of work into heat 
when $(\partial s/\partial\rho)_{T,\eta_e}<0$ as well as the loss of 
lepton pressure as a parcel of gas 
deleptonizes, which occurs as long as $(\partial Y_L/\partial\rho)_{T,\eta_e}
<0$.  Clearly, the growth rate given by eq. (\ref{growthrate}) 
can be thought of as two distinct stability criteria, one arising from 
the transport of heat and one due to the transport of lepton 
number.  Ignoring all geometric factors (which are not completely negligible)
the ``condition'' for radiative driving due to heat transfer is roughly
given by
\be
F_{Q}\gtrsim\frac{\omega\,{\tilde\omega}^2}{k^3v^2_A}\, P_b\simeq 
\frac{v^3_A}{c^3_i}\,P_b\,c_i,
\label{stability1}
\ee  
while the condition for radiative driving which derives from
lepton transport reads
\be
F_{L}\gtrsim\frac{\omega\,{\tilde\omega}^2}{k^3v^2_A} n\,Y_L\simeq 
\frac{v^3_A}{c^3_i}\, n\,Y_L\,c_i.
\label{stability2}
\ee
The above stability criteria apply for the slow mode (which always has a 
larger growth rate than the fast mode; BS03) in the limit that both $v^2_A
/c^2_i\lesssim 1$ and the background pressure support is 
primarily provided by the entropy of the baryons, which happens to be the  
case near the neutrinosphere.  The two conditions for instability can be 
reconciled by re-writing eq. (\ref{stability2}) as
\be
k_BTF_L\simeq Y_L\,F_Q\gtrsim\frac{v^3_A}{c^3_i}\,n\,k_BT\,Y_L\,
c_i\longrightarrow
F_Q\gtrsim\frac{v^3_A}{c^3_i}\,P_b\,c_i =\rho\,v^3_A.
\ee
Therefore, neutrino-driven magnetoacoustic instability can be 
summed up by this criterion for both transport processes.

Substituting input conditions appropriate for regions near the neutrinosphere
yields the instability condition
\be
{\rm\bf INSTABILITY}\,\,\,\,\,\,\,\, 
\left(\frac{L_{\nu}}{10^{53}{\rm erg\,s^{-1}}}\right)
\left(\frac{30\,{\rm km}}
{R_{\nu}}\right)^2\left(\frac{\rho_{\nu}}{10^{11}\,{\rm g\,cm^{-3}}}
\right)^{1/2}
\left(\frac{10^{15}\,{\rm G}}{B}\right)^3 \gtrsim 10^{-1}
\label{scalestability}
\ee  
where $L_{\nu},\,R_{\nu},\,{\rm and}\,\rho_{\nu}$ are the neutrino 
luminosity, radius of the neutrinosphere, and density at the 
neutrinosphere, respectively.  The stability criteria given 
by eq. (\ref{growthrate}) indicates that for field strengths at and above
the equipartition value such that $B^2/8\pi\sim P_b$,  the Alfv\'en speed 
in the above stability criteria can be replaced by the isothermal gas sound 
speed $c_i\sim (P_b/\rho)^{1/2}$.  The equipartition value of the magnetic 
field $B_{\rm eq}$ near the neutrinosphere is roughly given by
\be
B_{\rm eq}\simeq 2\times 10^{15}\left(\frac{\rho}{10^{11}{\rm\,g\,cm^{-3}}}
\right)^{1/2}\left(\frac{T}{4\,{\rm MeV}}\right)^{1/2}\,{\rm G},
\label{Beq}
\ee
which can be thought of as the magnetic field's limiting value for 
eq. (\ref{scalestability}) to be valid.  Deeper in the star, the degeneracy 
pressure of both the baryons and electrons overwhelms the contribution
from the baryons' entropy and the Fermi energies of both the electrons
and baryons, rather than temperature, will enter into eq. (\ref{Beq})
yielding a larger sound speed and thus a larger $B_{\rm eq}$. 

Though the stability criteria for both lepton and entropy driving 
are roughly equivalent, their respective growth and damping rates are not.  
From eq. (\ref{growthrate}), we see that the driving rate $\gamma_L$
originating from gradients in $\mu_{\nu_e}$, which are $\propto{\bf F_L}$, 
for the slow mode is approximated by
\be
\gamma_L\simeq \frac{1}{2}\frac{v_A}{c_i}\left(\frac{\partial P}
{\partial\mu_{\nu_e}}\right)_{\rho,T}\frac{\pi^2\kappa_0}{c_i\,{\mathcal 
K}}F_L&=&\frac{1}{2}\frac{v_A}{\rho\,c^2_i}\left(\frac{\partial P}
{\partial\mu_{\nu_e}}\right)_{\rho,T}|\bnabla\mu_{\nu_e}|
\nonumber\\
&\simeq&\frac{1}{2}\frac{v_A}{c_i}\frac{P_L}{P}\frac{g}{c_i}.
\label{leptongrowth}
\ee
The factor $P_L/P$, where $P_L$ is the lepton pressure, may not be entirely 
negligible near the neutrinosphere.  The driving rate due to gradients in 
temperature $\gamma_Q$, mediated by ${\bf F}_Q$, can be estimated 
in a similar fashion 
\be
\gamma_Q\simeq\frac{1}{2}\frac{v_A}{c_i}\left(\frac{\partial P}{\partial T}
\right)_{\rho,\mu_{\nu_e}}\frac{\kappa_{\rm eff}}{c_i\,{\mathcal K}
k^2_BT}\, F_Q&=&\frac{1}{2}\frac{v_A}{\rho c^2_i}\left(\frac{\partial P}
{\partial T}\right)_{\rho,\mu_{\nu_e}}|\bnabla T|\nonumber\\
&\simeq&\frac{1}{2}\frac{v_A}{c_i}\frac{P_b}{P}\frac{g}{c_i},
\label{heatgrowth}
\ee
where $P_b$ is the baryon pressure.  Near the neutrinosphere, $P_b\sim P$,
but $P_L$ is certainly non-negligible.  Noting that $P_L\simeq
P_e\sim(\eta_e/4)\,Y_e\,P_b$ indicates that if radiative driving is
allowed to occur at relatively large depths, where the degenerate lepton
pressure becomes formidable, then the neutrino bubble instability results
primarily from lepton, rather than heat, transport.  

In terms of equilibrium parameters near the neutrinosphere, the combined 
neutrino bubble growth rate $\gamma_{\rm NB}\equiv\gamma_L+\gamma_Q$ 
is approximately 
\be
\gamma_{\rm NB}\simeq 10^{3}\left(\frac{B}{10^{15}\,{\rm G}}\right)
\left(\frac{4\,{\rm MeV}}{T}\right)\left(\frac{10^{11}\,{\rm
g\,cm^{-3}}}{\rho}\right)^{1/2}
\left(\frac{g}{10^{13}\,{\rm cm\,s^{-2}}}\right) 
s^{-1}. 
\label{roughgrowth}
\ee  
As in the case of the instability criteria eq. (\ref{scalestability}),
the value of the magnetic field $B$ in the slow mode growth
rate possesses an upper limit whose value is given by eq. (\ref{Beq}). 
   
Let us summarize the results of this section.  Near the neutrinosphere, 
the condition for local driving due to the presence of ${\bf F_Q}$ and 
${\bf F_L}$ are roughly equivalent.  This is true primarily because
the neutrino chemical potential $\mu_{\nu_e}\sim k_BT$, which implies that the 
flux of leptons directly maps to the flux of heat modulo a factor that is
$\sim Y_L$.  Neutrino bubble instabilities can grow on millisecond 
timescales for strong magnetic field strengths, with growth times
$\gtrsim 1\,{\rm ms}$ for equipartition magnetic fields.  Driving due to 
the presence of ${\bf F_Q}$ mostly likely dominates the 
driving near the neutrinosphere over the radiative driving provided 
by ${\bf F_L}$.

\section{Monte Carlo Calculation}\label{s:montecarlo}

Assuming that radiative driving occurs in the upper layers of a PNS,
neutrino bubbles may then lead to substantial luminosity enhancements and 
large density fluctuations in strongly magnetized regions.  In order to 
make quantitative statements regarding the non-linear outcome of the 
neutrino bubble instability, a theory
of radiation-driven magnetoacoustic turbulence is required.  To our 
knowledge, such a theory unfortunately does not exist.  We speculate
how the saturation amplitude may change for different values of the 
local magnetic field strength in \S\ref{ss:saturation} under the 
assumption that significant non-linearities occur even for gas pressure
dominated equilibria.  

In order to proceed, we approximate the effects of the neutrino bubble 
instability by artificially altering the neutrino opacity in a radial and 
angular cut surrounding the neutrinosphere at early times after bounce.
By doing so, we capture the cumulative effect that we expect neutrino bubbles 
produce on the surface of a PNS.  From this deformed stellar 
background, the emitted neutrino flux is simulated  
via a Monte Carlo calculation so that the level of asymmetry in the 
outgoing radiation beam can be estimated.  The angular profile of the neutrino 
flux allows us to further estimate the magnitude of the instantaneous kick 
as well as the gravitational wave strain and r-process efficiency.

\subsection{Numerical Inputs and Techniques}

There are three ingredients that go into our  
numerical model of the neutrino bubble powered mechanism:  atmospheric 
structure, parameterization of the neutrino bubble instability, and
radiative transfer.  We describe each of them below.

\subsubsection{Proto-neutron star structure}

The PNS that we use in our simulation was born from a 15$M_{\odot}$ 
progenitor (s15s7b - Woosley \& Weaver 1995).   Collapse, bounce, and 
early contraction are evolved up to a time of 130 ms after bounce
using a 1-dimensional core-collapse code (Herant et al. 1994; Fryer 
et al. 1999).  To this point, we assume that the evolution 
of the star is spherically symmetric and neutrino bubble oscillations 
have not yet begun.

\subsubsection{Neutrino Bubbles as a modified opacity}

In our stellar model, the onset of neutrino bubble instability 
introduces deviations from spherical symmetry.  As discussed in 
\S\ref{ss:magconv}, neutrino bubbles are driven in starspots 
above the convection zone, but below the neutrinosphere.  Effectively, 
one can model the global radiation asymmetry due to potential luminosity
enhancements in starspots by reducing the opacity in those regions
while the star's hydrostatic structure remains the same.  By doing so, 
optically thick neutrinos
preferentially leak out towards these starspots, producing a cumulative 
asymmetry in the neutrino beam, which emulates the effect of the 
acoustic instability.  If in {\it reality}, the 
neutrino bubble instability behaves as the photon bubble instability in the
non-linear regime,
then large density fluctuations occur and the radiation preferentially
leaks out of the rarefied regions, locally enhancing the 
outgoing flux (Begelman 2001, Turner et al. 2004).
\footnote{Drawing conclusions regarding neutrino bubble saturation in 
PNSs from work done on photon bubbles in luminous accretion 
flows is dangerous.  Those authors considered radiation pressure
dominated media with superstrong magnetic fields.  For PNSs, the magnetic
field is not superstrong and the radiation field is significantly 
sub-Eddington in neutrinos.} 

The angular extent of a starspot is parameterized by a covering 
fraction ${\mathcal C}$ and values for ${\mathcal C}$ are taken to be from 
0.01-0.1.  The radial extent for a neutrino bubble-active starspot is 
parameterized by an inner- and outer-radius ${\mathcal R}_{in}$ and 
${\mathcal R}_{out}$, respectively.  A largely phenomenological justification 
for these values was given in \S\ref{ss:magconv}.
Throughout the neutrino bubble-active 
region, we reduce the neutrino opacity by a factor {\it f}.  
This reduction in opacity crudely characterizes the potentially
multi-dimensional saturated state of the neutrino bubble instability 
into a single parameter.  A list of values for ${\mathcal C}$, 
${\mathcal R}_{in}$, ${\mathcal R}_{out}$, and $f$ that are employed in our 
calculations are given in Table \ref{t:1}.

\subsubsection{Neutrino Transport}

We simulate the transport of neutrinos through our neutrino
bubble-deformed PNS 130 ms after bounce via the Maverick Monte Carlo
transport code (Hungerford et al. 2003).  In our calculations, we
make several simplifying assumptions.  We only model the electron
neutrinos $\nu_e$, while assuming that the asymmetry they
experience is representative of all neutrino species.  We also assume
that the asymmetry in the neutrino emission remains the same for the
entire phase of neutron star cooling, which will inevitably lead to 
an overestimate of the net asymmetry.  With these caveats, we can then
calculate the total momentum kick and gravitational wave signal
based on the total energy emitted in neutrinos for our neutron star
(for our model, the total energy emitted is roughly 
$5\times10^{53}$\,erg).

Neutrino/matter interactions are calculated using a simplified opacity
given by (Janka 2001)

\begin{equation}
\kappa_{a} \approx {\frac{3 \alpha^{2} + 1}{4}}\frac{\sigma_0}{m_n}
\frac{\left<\epsilon^2_{\nu_e}\right>}{\left(m_ec^2\right)^2}\,
Y_{n}
\label{Janka1}
\end{equation}

\noindent and

\begin{equation}
\kappa_{sc} \approx {\frac{5 \alpha^{2} + 1}{24}}\frac{\sigma_0}{m_n}
\frac{\left<\epsilon^2_{\nu_e}\right>}{\left(m_ec^2\right)^2}
(Y_n+Y_p).
\label{Janka2}
\end{equation}

Here, $\sigma_0 = 1.76 \times 10^{-44}$~cm$^{2}$ is the characteristic 
weak interaction cross section and $\alpha$ is the vector coupling constant
in the vacuum. The absorption opacity $\kappa_a$ is due to electron 
capture and the scattering opacity $\kappa_{sc}$ results from elastic
scattering off of nucleons.  Neutrino annihilation was not included as 
an opacity source since the annihilation cross section 
becomes vanishingly small near the neutrinosphere, when compared to 
other sources of opacity.  This follows from the fact that neutrinos 
propagate almost purely radially.  

Neutrino emission rates from electron capture  and
electron-positron annihilations are taken from the temperature and
density dependent functions given by Colgate, Herant \& Benz (1993)
and the energy distribution for the emitted
neutrinos is taken from a simplified functional fit to the Fermi-Dirac
distribution, as described by Keil, Raffelt \& Janka (2003).  Emitted 
neutrino energies were sampled from their Fermi sea with a two 
step (inversion followed by rejection) sampling technique.
\begin{deluxetable}{cccccc}
\center
\tablewidth{0pt}
\tablecaption{Monte Carlo Simulations}
\tablehead{
\colhead{$\mathcal{C}$} &
\colhead{${\mathit f}$} &
\colhead{$\mathcal{R}_{out}$}&  
\colhead{$\mathcal{R}_{in}$}&
\colhead{$V_{kick}$} & 
\colhead{$<h_{TT}>_e$}\\
\colhead{}  & 
\colhead{} & 
\colhead{(km)} & 
\colhead{(km)} & 
\colhead{(km/s)} & 
\colhead{(\@ 10~kpc)}}
\startdata
0.0  & 1    & --- & -- &  100 &  1.0E-20    \\
0.1  & 2    &  90 & 50 & 1675 &  1.3E-19 \\
0.1  & 2    &  90 & 60 &  965 & \\
0.1  & 1.33 &  90 & 50 &  700 & \\
0.1  & 1.33 &  90 & 60 &  520 & \\
0.1  & 1.11 &  90 & 60 &  350 & \\
0.1  & 1.11 &  90 & 60 &  220 & \\
0.05 & 2    &  90 & 50 &  890 & \\
0.05 & 2    &  90 & 60 &  530 & 5.E-20 \\
0.01 & 2    &  90 & 50 &  175 & \\
0.01 & 2    &  90 & 60 &  120 & \\
0.1  & 2    & 120 & 50 & 2350 & \\
0.1  & 2    & 120 & 60 & 1500 & \\
0.1  & 1.33 & 120 & 50 &  950 & \\
0.1  & 1.33 & 120 & 60 &  700 & \\
0.05 & 2    & 120 & 50 & 1250 & \\
0.05 & 2    & 120 & 60 &  800 &   \\
0.01 & 2    & 120 & 50 &  250 &    \\
0.01 & 2    & 120 & 60 &  200 &    \\
\enddata
\label{t:1}
\end{deluxetable}

\begin{figure}[t!]
\label{deposition}
\begin{center}
\epsfxsize=3.5truein 
\epsffile{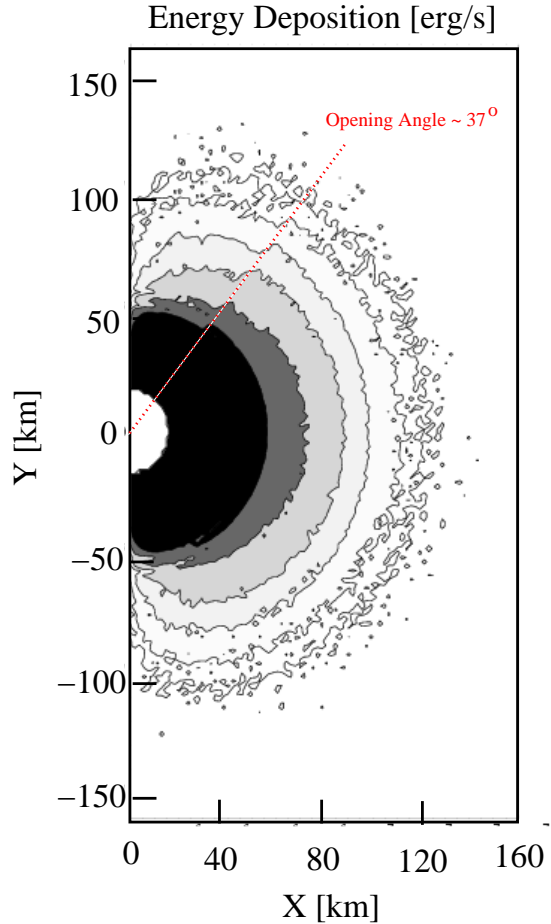}
\caption{Neutrino energy deposition contours for
$\mathcal{C}=0.1$, ${\it f}=2$, $\mathcal{R}_{out}=120$~km,
$\mathcal{R}_{in}=50$~km.  Contour levels plotted are 0.06, 0.2, 0.8, 
2.2, 7 and 27 in units of 10$^{50} {\rm ergs\, s^{-1}}$.  Note that the 
contours move inward slightly over the cone of lower opacity (cone 
opening angle is roughly 37$^{\circ}$ for
$\mathcal{C}=0.1$).   This slight deviation from spherical symmetry 
will not likely
cause strong variations in the PNS wind.  The decrease in deposited 
energy right
along the symmetry axis is the result of insufficient sampling due to 
the smaller
volume at small radii.}
\end{center}
\end{figure}

Using appropriate Monte Carlo estimators, we track the spatial distrubition
of neutrino energy deposition, luminosity with respect to polar angle,
 and the net neutrino momentum.   Neutrino deposition is tallied in a 
2D cylindrical grid with 88 radial and 176 vertical bins.  The Monte Carlo
estimator for this quantity is the sum of packet luminosity for all packets
absorbed at cylindrical coordinate {\it (r,z)}.  The neutrino packets have
uniform luminosity weight, so the Monte Carlo noise associated with
the calculated deposition is proportional to the packet luminosity times
$\sqrt{N(r,z)}$, where $N(r,z)$ is the number of packets absorbed at
cylindrical coordinate {\it (r,z)}.  A typical value for the noise is
roughly 0.003\%, though along the polar
axis the noise is significantly higher as the sampled cylindrical
volume drops to zero at small radii.  This purely numerical effect can be 
seen as dips in the deposition profile along the poles (see Figure
3).  Note that 
numerical noise is given in terms of energy per unit time since we are 
only considering a short time-slice in the evolution of the PNS.  

The angular distribution of the neutrino luminosity is similarly
calculated by summing the
luminosity of escaping neutrino packets into 11 angular bins along the
polar direction.  For a given bin direction $\theta$, packets escaping
with direction $\theta \pm \Delta \theta$ are counted in the
sum (where $\Delta \theta =$ 2.5$^{\circ}$).  As before, the noise goes as
the packet luminosity times $\sqrt{N(\theta)}$, where $N(\theta)$ is the 
number of packets in angular bin $\theta$, which works out to be roughly 
0.01\% (see Figure 3).  The Monte Carlo 
estimator for net neutrino momentum is given, in component
form ($p_x$, $p_y$ and $p_z$), by
\be
p_x = \sum_0^{N_{esc}} \frac{L_{pckt}}{c} \times\mu_x\,;\,\,\,\,\,\,\,
p_y = \sum_0^{N_{esc}} \frac{L_{pckt}}{c} \times\mu_y\,;\,\,\,\,\,\,\,
p_z =\sum_0^{N_{esc}} \frac{L_{pckt}}{c} \times\mu_z,
\ee
where $N_{esc}$ is the number of escaped packets, $L_{pckt}$ is the weight
of the packet in units of luminosity, $c$ is the speed of light and
$\mu_x, \mu_y, \mu_z$ are the direction cosines with respect to the coordinate
axes.  This is the most well determined quantity from these simulations, as
it is tabulated from all escaping neutrino packets ($N_{esc}
\approx 2 \times 10^6$).  The uncertainty in the net neutrino momentum
per unit time resulting from angular discretization is of order 
10$^{40}{\rm g cm s^{-2}}$.  As the first line of Table 1 shows, 
numerical noise alone is responsible 
for momentum asymmetries in our calculations, leading to velocity 
kicks $\sim 100\,{\rm km\,s^{-1}}$ if this artificial asymmetry where 
to continuously discharge its momentum onto the star for $\sim 1$s.

\subsection{Numerical Results}

Now we are in the position to provide estimates of the kick imparted, 
r-process yields, and gravitational wave strains which may result
from the neutrino bubble instability.  Values for these observable
quantities are provided in Table \ref{t:1}.

Our work focuses on the origin of neutron star kicks, which is the 
most dramatic effect of neutrino bubble instability on the 
core-collapse enviroment.  The total momentum
carried away by the neutrinos in the collapse is equal to the total
neutrino energy emitted divided by the speed of light.  The asymmetry
in neutrino emission carries away a net linear momentum.  The total 
asymmetry is measured at infinity rather than at the neutrinosphere.  
Thus, the optically thin matter above the region of neutrino bubble
instability reduces the asymmetry by a multiplicative factor 
$\sim e^{\tau_{\nu}}$ and 
most likely leads to an underestimate of the momentum kick.  As expected, 
Table \ref{t:1} shows that neutrino bubbles may drive kicks in excess of 
$1000\,{\rm km\,s^{-1}}$.  We assume that the kick is discharged 
for $\sim 1$s, which can be thought of as the best case scenario.  Further,
we are implicitly assuming that the spin and the starspot are exactly 
aligned.  This assumption overestimates the kick by the value of the 
cosine of the angle between the spin axis and the starspot.

In order for neutrino
bubbles to affect the r-process or the supernova explosion, it
need not alter the total energy deposition of neutrinos dramatically,
but it must induce an asymmetry in this deposition.  Our simulations
show that this instability produces only a small asymmetry, and the
total heating from the proto-neutron star is essentially spherically
symmetric, even for our most extreme case (see Figure 3).

However, these small asymmetries may lead to large differences in the
gravitational wave output and spectrum.  The gravitational wave amplitude
($h^{\rm TT}_{xx}$) for an observer orthogonal to the symmetry axis for
asymmetric neutrino emission is given by (M\"uller \& Janka 1997;
Fryer et al. 2004):
\begin{equation}
(h^{\rm TT}_{xx})_{\rm pole} = \frac{2G}{c^4 R} \sum \Delta t
\sum_{p=1}^{N} (1+z/r) (2x^2/r^2-1) \Delta L_{\nu}.
\end{equation}
The neutrino luminosity, $\Delta L_{\nu}$, at a position {\it x, y, z}
($r=(x^2+y^2+z^2)^{1/2}$) over a given time ($\Delta t$) must be summed
over all time to give the total wave amplitude.  For our suite of
models, the gravitational wave amplitude (assuming the asymmetry
persists for 1\,s) is sufficiently
high to be detectable by advanced LIGO observing a Galactic
supernova.\footnote{Depending upon the duration of the explosion, the
neutrino luminosity can vary considerably over this 1\,s time period
and it is possible that the gravitational wave amplitude may be even
higher (perhaps lower) than this predicted value, but this value
gives a rough estimate of the expected signal.}
Although this signal will be very different than the gravitational
wave signal from rotating supernovae, it will not be dramatically
different from other asymmetries in supernovae (e.g. asymmetric
collapse - Burrows \& Hayes 1996; Fryer 2004).


\section{Uncertainties, Comparison with Other Kick Mechanisms, and
Future Work}\label{s:conclusions}

The neutrino bubble powered kick mechanism presented here 
is subject to several uncertainties, many of which have not 
been mentioned thus far.  In what follows, we assess its major 
weaknesses and by doing so, provide motivation for future work.       

\subsection{Global Stability}\label{ss:global}

We have focused on the physics of {\it local} magnetoacoustic
stability in the presence of a stratified degenerate optically thick 
radiation field.  We can not stress enough, that drawing conclusions regarding 
stability from a local linear analysis is dangerous at best.  In order to 
confirm that neutrino bubbles are important in PNSs, a {\it global} analysis 
must be performed using a full non-adiabatic treatment since the 
region of greatest driving lies near the neutrinosphere, where the 
radiative diffusion time is short in comparison to the acoustic crossing 
time.  Our kick model requires a magnetic field that perhaps originates
from vigorous convective motion beneath the upper radiative layer.  
Turbulent motions almost certainly reduce the driving rate of a given 
trapped wave (Goldreich \& Keeley 1977).  However, even if the interaction 
between a mode and turbulence is accurately quantified, 
calculating the cumulative damping rate throughout the run of the star 
requires knowledge of the mode's profile as a function of depth.     

\subsection{Saturation Amplitude}\label{ss:saturation}

Assuming that one finds trapped neutrino bubble oscillations in a global 
eigenmode calculation, computing the saturation amplitude is a much more 
difficult task since a fully
non-linear degenerate radiation magnetohydrodynamic analysis is required.
But, from a computational expense point of view, the situation may not be 
as costly as in the case of main sequence stars and the 
accretion disks that are thought to power
active galactic nuclei and X-ray binaries.  Since 
the sound speed is relatively close to the speed of light in PNSs, the 
ratio of the sound crossing time to the radiation diffusion time near
the surface of last scattering is only a factor of $\sim 30$
or so.  For the A-star discussed in \S \ref{s:mechanism}, this ratio 
is $\sim 10^4$.  


We expect that the saturation amplitude becomes smaller for 
decreasing values of the magnetic field strength.  If the saturation 
amplitude depends on the ratio of equilibrium
magnetic to gas pressure evaluated at the neutrinosphere, then large flux 
enhancements would not occur for local
field strengths $< 10^{15}$ G.  However, if the saturation amplitude varies
with respect to the ratio of the Alfv\'en speed $v_A$ to the gas sound 
speed $c_i$, then perhaps field strengths of order $\sim$ a few times
$10^{14}$ G may lead to a saturation amplitude of order $v_A/c_i\sim 1/3$.
Of course, the entire kick model relies on the fact that significant 
luminosity enhancements are reached for thermally- 
and chemically-locked perturbations in gas pressure dominated media.  
The results of \S\S \ref{s:thermodynamics}, \ref{s:mechanics}, and of BS03
inform us that dynamical growth rates can be achieved for wavelengths of 
order the gas pressure scale height (where our WKB analysis breaks down).  
Therefore, the energy injected into the mode by the instability is 
comparable to the kinetic energy of the mode itself.  Whether or not 
this leads to order unity saturation amplitudes is a question that can 
only be answered with confidence by a careful non-linear analysis.

\subsection{Magnetic Field Structure}\label{fieldstructure}

The neutrino bubble instability can not power a kick 
unless strong magnetic structures, with 
length scales significantly larger than the gas pressure scale height
at the neutrinosphere, are maintained for hundreds of convective 
overturn times.  
Reasons for believing that such fields exist in young PNSs were
given in \S\ref{ss:magconv}.  Our arguments regarding magnetic field 
structure were primarily based on phenomenological arguments derived from 
the Sun.  However, even after the discovery of the solar tachocline (Goode et
al. 1991) and the understanding of its importance with respect to magnetic
field generation (Spiegel \& Zahn 1992), the evolution of the Sun's surface
magnetic field throughout the solar convection zone evades a coherent 
theoretical picture.  Numerical simulations of PNS magnetic field 
structure may prove to be more fruitful than those of their main sequence 
counterparts since a PNS convection zone encompasses relatively few gas 
pressure scale heights in comparison to say, the solar convection zone.

\subsection{Handling of the Neutrino Transport}
 
We have ignored the compositional dependence of the neutrino opacities
in our linear analysis.  
Not only does this change the numerical value of the opacity, but more 
importantly, we may have overlooked certain instability phenomena as well.  
In general, the neutrino 
absorption opacity depends on the proton fraction (see eq. (\ref{Janka1}) and 
Janka 2001).  Since the proton fraction $Y_p$ depends on fluid's density,
the opacity perturbation $\delta\kappa$ varies with the density perturbation 
$\delta\rho$ implying that standard hydrodynamic sound waves which display
some degree of radial propagation can be driven unstable by the presence 
of the radiative fluxes (Unno et al. 1989, Glatzel 1994,
 BS03).  Therefore, when the compositional variation of the absorption 
opacity is taken into account, $\kappa$-mechanism and strange mode 
phenomena may ensue on the entire surface of the proto neutron star i.e.,
radiative driving occurs in un-magnetized regions as well.  
If thermal locking holds, the growth rate near the neutrinosphere
from local radiative driving is comparable to that given by the upper limit
of  eq. (\ref{roughgrowth}).  Certainly, if the $\kappa$-mechanism or strange 
mode instability occurs at early times, luminosity perturbations of only 
$\sim 10\%$ could play an important role with respect to shock ejection
and the overall explosion mechanism.  
Fully understanding how this different instability phenomenon affects
the neutrino bubble kick mechanism requires answering the questions
posed by \S\S \ref{ss:global}-\ref{fieldstructure}.

\subsection{Comparison with Other Kick Mechanisms}

If these theoretical hurdles are overcome, a neutrino bubble powered
kick mechanism possesses attractive differences when compared to other
mechanisms.  In order to obtain spin-kick alignment, as is seen in the
Crab and Vela systems, the impulse must be imparted on timescales
significantly in excess of the rotation period (Lai, Cordes, \&
Chernoff 2001).  In ``mass-rocket'' models, a PNS receives various
impulses with durations of order $\sim$ a few milliseconds each
(Burrows \& Hayes 1996; Lai \& Goldreich 2000; Fryer 2004) .
Therefore, spin-kick alignment most likely does not follow from the
hydrodynamic mass-rocket mechanism unless the initial rotation period
is $\lesssim$ a millisecond.  However, this fact only rules out the 
mass-rocket model for the Crab and Vela systems, which may not be  
representative of the Galactic pulsar distribution.   Fryer (2004) 
argues that the mass-rocket works only for rapid explosions 
so that the convective motions in the gain region do not have enough time 
to damp out the mean motion of the PNS.  Therefore, mass-rocket kick 
models are heavily prejudiced towards neutron stars born from progenitors
with mass $<12M_{\odot}$.

Recently, another hydrodynamic powered kick model that depends on delayed 
explosions triggering global Rayleigh-Taylor instability
has been further elaborated by way of detailed numerical simulation 
(Scheck et al. 2004).  Their model
requires the stellar neutrino emission to be somewhat uniformly discharged over
the first second so that the refreshment epoch is prolonged.  This
allows global Rayleigh-Taylor type instability to develop beneath
the shock front since the density and pressure gradients lie in the
opposite direction across the shock (Thompson 2000, Herant 1995).  
A prolonged refreshment is necessary in order to allow the lowest order 
($l=1,2$) modes to dominate the non-linear phase.  This mechanism does not 
lead to spin-kick alignment and most likely requires weak explosions. 

The parity violation model of Arras \& Lai (1999a, 1999b) naturally
produces spin-kick alignment as do all neutrino-powered kick mechanisms.  
This results from the basic fact that the Kelvin-Helmholtz time, which 
is the discharge timescale for neutrino powered kicks, is most likely much 
longer than the rotation period.  Also, neutrino-driven kick mechanisms are 
not subject to convective damping in the gain region as in the case of the
mass-rocket mechanism since the the shell of accreting material 
beneath the outgoing shock can easily adjust its center of mass to 
that of the kicked PNS.  As previously mentioned, 
the parity-violation mechanism requires global magnetic flux densities 
$\sim 10^{16}$ G, a large value even for magnetars.  Say for example, that 
the neutrino bubble powered kick mechanism requires flux densities of 
comparable strength in order to produce order unity luminosity 
enhancements over a  starspot.  Then, such a starspot would only need to 
cover $\sim 10\%$ of a hemisphere.  Thus, the total amount of flux 
necessary for the neutrino bubble mechanism, for this rather extreme case,
is $\sim 10\%$ of that required for the parity violation model.  In terms of
the overall magnetic energy budget, the neutrino bubble mechanism requires
$\sim 1\%$ of the total magnetic energy with respect to the parity violation 
model.  

There may not be a single explanation for neutron star kicks.
Perhaps all, some, or none of the kick mechanisms mentioned here
(including ours) work for different classes of progenitors.  The best
way to differentiate between the various kick mechanisms is to
constrain models with observations and analysis of the Galactic pulsar
and neutron star population.  For example, determining the fraction of
systems which possess spin-kick alignment, quantifying the robustness of the
inferred bimodal distribution as well as the correlation between kick 
velocity and magnetic field strength will help constrain the applicability
of these various mechanisms.

\acknowledgements{We thank P. Arras , L. Bildsten, A. Burrows, P. Chang, 
B. Pacy\'nski, R. Rosner, C. Thompson, and N. Turner for enlightening  
conversations regarding a variety of topics.  This work was supported 
by Los Alamos National Laboratory/UCSB CARE grant SBB-001A, and
NASA grant NAG5-13228.  A portion of this
work was also funded under the auspices of the U.S. Department of 
Energy through contract W-7405-ENG-36 to Los Alamos National Laboratory.  
The simulations were conducted on Space Simulator at Los Alamos National 
Laboratory. A.S. acknowledges financial support from a Hubble Fellowship 
administered by the Space Telescope Science Institute.}

\appendix

\section{Appendix A: Thermodynamic Quantities and Relations}\label{a:a}

Here, we derive the expressions for the neutrino energy 
and lepton flux
as well as other important thermodynamic quantities.  Much of this 
Appendix is covered by Bludman \& Van Riper (1978)
as well as Lai \& Qian (1998).  

All spin 1/2 particle species are assumed to possess a Fermi-Dirac
distribution $f$ where 
\be
f_i=\frac{1}{e^{(E-\mu_i)/k_BT}+1}
\ee
for the $i^{\rm th}$ species.  Thus, any given particle 
distribution is parameterized solely by a temperature and 
chemical potential, both of which vary as a function of
depth.  In the diffusion approximation the energy density, number
density, energy flux, and particle flux are given by 
\be
U=\int dE\,  U_E=\int dE\frac{4\pi}{h^3c^3}\,E^3\,f,\,\,\,\,\,\,\,
n=\int dE\,\frac{U_E}{E},\,\,\,\,\,\,\,\,\,
\,\,\,\,\,\,\,\,\,{\bf F}=-\frac{c}{3\rho}
\int dE\frac{1}{\kappa(E)}\bnabla U_E,\,\,\,\,\,\,\,\,\,
{\rm and}\,\,\,\,\,\,\,\, {\bf H}=-\frac{c}{3\rho}\int dE
\frac{1}{\kappa(E)}\bnabla\frac{U_E}{E}
\ee 
respectively.  For sake of simplicity, we assume that $\kappa(E)$ 
depends solely on energy as given by eq. (\ref{opacity}) and is the 
sum of both the scattering and absorption opacities.  Integrating 
over energy we obtain
\be
U=\frac{k^4_BT^4}{2\pi^2\hbar^3c^3}F_3\left(\eta\right),\,\,\,\,\,\,\,
\,\,n=\frac{k^3_BT^3}{2\pi^2\hbar^3c^3}F_2\left(\eta\right),\,\,\,\,\,\,\,
{\bf F}=-\frac{E^2_0k^2_B}{6\pi^2\kappa_0
\rho\hbar^2c^2}\bnabla\left[T^2F_1(\eta)\right],\,\,\,\,\,\,\,\,
{\rm and}\,\,\,\,\,\,\,{\bf H}=-\frac{E^2_0k_B}{6\pi^2\kappa_0
\rho\hbar^2c^2}\bnabla\left[T\,F_0(\eta)\right]
\ee
where $F_n(\eta)$ is the standard Fermi-Dirac integral given by
\be
F_n(\eta)=\int^{\infty}_0dx\,\frac{x^n}{e^{x-\eta}+1}.
\ee
The chemical potential of the electron-type neutrinos obey the 
relation $\mu_{\nu_e}=-\mu_{{\bar\nu_e}}$ which allows us to write
\be
U_{\nu_e}+U_{{\bar\nu_e}}=\frac{7\pi^2k^4_BT^4}{120\hbar^3
c^3}\left[1+\frac{30}{7}\left(\frac{\eta_{\nu_e}}{\pi}\right)^2+
\frac{15}{7}\left(\frac{\eta_{\nu_e}}{\pi}\right)^4\right],
\ee
\be
Y_{\nu_e}=\frac{n_{\nu_e}-n_{\bar{\nu_e}}}{n}={\kb^3T^3\over6n
\hbar^3c^3}
\eta_{\nu_e}\left(1+{\eta_{\nu_e}^2\over\pi^2}\right),
\ee
\be
{\bf F}_{\nu_e}+{\bf F}_{{\bar{\nu_e}}}=-\frac{E^2_0k^2_B}{6\pi^2
\kappa_0\rho\hbar^3c^2}\bnabla\left[T^2\left(\frac{\eta^2_{\nu_e}}{2}
+\frac{\pi^2}{6}\right)\right],
\ee 
and
\be
{\bf F_L}={\bf H}_{\nu_e}-{\bf H}_{{\bar\nu_e}}=
-\frac{E^2_0k_B}{6\pi^2\kappa_0
\rho\hbar^2c^2}\bnabla\left[T\eta_{\nu_e}\right].
\ee
The lepton flux, ${\bf F_L}$, is solely carried by the electron-
type neutrinos due to their macroscopic mean free paths and because
the $X-$type neutrinos are produced only from pair annihilation 
processes.  The above relations were obtained with the help of the following
expressions (Bludman \& Van Riper 1978)
\be
F_0(\eta)-F_0(-\eta)&=&\eta\nonumber\\
F_1(\eta)+F_1(-\eta)&=&\frac{\eta^2}{2}+\frac{\pi^2}{6}\nonumber\\
F_2(\eta)-F_2(-\eta)&=&\frac{\eta}{3}\left[\eta^2+\pi^2\right]\nonumber\\
F_3(\eta)+F_3(-\eta)&=&\frac{7\pi^4}{60}+\frac{1}{2}\eta^2\left[
\pi^2+\frac{1}{2}\eta^2\right].
\ee   
The number densities, energy densities, and fluxes for all other 
fermionic species (pairs and $X-$type neutrinos) can be quickly obtained 
repeating the procedure outlined above.  

\section{Appendix B: Conditions for Silk Damping, Diffusive Heating, and 
Damping from Lepton Pressure Loss}\label{a:b}

In order to see if diffusive heating occurs at all for short wave-length 
compressible waves, the ``Silk determinant'' must be greater than zero
i.e., 
\be
n\,T\left(\frac{\partial s}{\partial\rho}\right)_{T,\mu_{\nu_e}}
&=&\left(\frac{\partial U}{\partial\rho}\right)_{T,\mu_e}-
\frac{U+P}{\rho}-\mu_{\nu_e}n\left(\frac{\partial Y_L}
{\partial\rho}\right)_{T,\mu_{\nu_e}}>0.  
\label{Silk}
\ee
In order to evaluate this expression, we need to specify an equation 
of state.  Photons and leptons obey relativistic equations of state
for bosons and fermions, respectively.  However, the quantum 
mechanical behavior for the non-relativistic baryons varies with 
increasing depth.  In the inner and outer core, the baryons are mildly 
degenerate, while near the neutrinosphere baryons behave as a
classical ideal gas ($P\propto n\,T$).

It is convenient to define a fiducial Fermi energy for the 
non-relativistic baryons
\be
\epsilon_{F}\equiv\frac{\hbar^2}{2m_n}p^2_F=\frac{\hbar^2}{2m_n}
\left(3\pi^2 n\right)^{2/3}\simeq 124 \,\rho^{2/3}_{15}{\rm MeV}
\ee
such that the Fermi energy and and zero-point pressure for the neutrons 
and protons can be conveniently written as
\be
\epsilon_{n,p}\equiv Y^{2/3}_{n,p}\epsilon_F\,\,\,\,\,\,{\rm and}\,
\,\,\,\,\,\,P^{(0)}_{n,p}=\frac{2}{3}U^{(0)}_{n,p}= \frac{2}{5}Y^{5/3}
_{n,p\,}n\,\epsilon_{F} 
\label{Pfermi}
\ee
respectively.  At early times in the mantle and near the 
neutrinosphere, $\epsilon_F\simeq 6\,\rho^{2/3}_{12}\,{\rm MeV}$ while 
$k_BT\simeq 10-30\,{\rm MeV}$ such that the contribution to the
baryon pressure arising from degeneracy may be neglected at first 
approximation.  Deep in the core below the mantle, the degeneracy 
pressure of the baryons is substantial such that $\epsilon_{n,P}
\sim Y^{2/3}_{n,p} 60\,{\rm MeV}$ is significantly larger than 
$k_BT\simeq 15-20\,{\rm Mev}$ (Pons et al. 1999).  Their pressure 
is approximately given by
\be
P_{n,p}\simeq P^{(0)}_{n,p}\left[1+\frac{5\pi^2}{12\,Y^{4/3}_{n,p}}
\left(\frac{k_BT}{\epsilon_{F}}\right)^2\right]
\ee
where we have kept the substantial finite temperature contribution for 
each baryonic species.  The only species significantly contributing 
to the 
first term in eq. (\ref{Silk}) are the neutrons and electrons where
\be
\left(\frac{\partial U_n}{\partial\rho}\right)_{T,\mu_e}=
\left(\frac{\partial U_n}{\partial\rho}\right)_{T,\mu_e,\mu_{\nu_e}}
+\left(\frac{\partial\mu_e}{\partial\rho}\right)_{T,\mu_{\nu_e}}\left(
\frac{\partial U_n}{\partial\mu_e}\right)_{\rho,T,\mu_{\nu_e}}\simeq
\frac{5}{3\,Y_n}\frac{U^{(0)}_n}{\rho}\left[1+\frac{5\pi^2}
{12\,Y^{4/3}_n}\left(\frac{k_BT}{\epsilon_F}\right)^2\right]
-\frac{4}{3\,Y_n}\frac{U^{(0)}_n}{\rho}\frac{5\pi^2}{12\,Y^{4/3}_n}
\left(\frac{k_BT}{\epsilon_F}\right)^2
\ee 
\be
\left(\frac{\partial U_e}{\partial\rho}\right)_{T,\mu_{\nu_e}}=
\left(\frac{\partial\mu_e}{\partial\rho}\right)_{T,\mu_{\nu_e}}\left(
\frac{\partial U_e}{\partial\mu_e}\right)_{\rho,T,\mu_{\nu_e}}\simeq
\frac{3\,Y_e}{Y_n}\frac{k_BT}{m_n}.
\label{silkelec}
\ee
Terms ${\mathcal O}(k_BT/\mu_e)$ have been dropped for 
the sake of simplicity, which is why the proton contribution has been 
neglected.  This is not a bad approximation during early times 
throughout the core where $\mu_e/k_BT\sim 10-20$ (Burrows \& Lattimer 
1986, Pons et al. 1999).  

The last term in eq. (\ref{Silk}), which is largely responsible for 
diffusive heating of compressible perturbations is given by
\be
-\mu_{\nu_e}\,n\,\left(\frac{\partial Y_L}{\partial\rho}\right)_{T,
\mu_{\nu_e}}
=-\mu_{\nu_e}\,n\left[\left(\frac{\partial Y_L}{\partial\rho}\right)_{T,
,\mu_e,\mu_{\nu_e}}+\left(\frac{\partial\mu_e}{\partial\rho}\right)_{T,
\mu_{\nu_e}}\left(\frac{\partial Y_L}{\partial\mu_e}\right)_{\rho,
T,\mu_{\nu_e}}\right]\simeq+\frac{\mu_{\nu_e}}{m_n}Y_L.
\ee      
Thus, to ${\mathcal O}(k_BT/\mu_e)$ the Silk determinant is given 
by
\be
n\,T\left(\frac{\partial s}{\partial\rho}\right)_{T,\mu_{\nu_e}}
&\simeq&
\frac{5\,Y_p}{3\,Y_n}\frac{U^{(0)}_n}{\rho}\left[1+\frac{5\pi^2}
{12\,Y^{4/3}_n}\left(\frac{k_BT}{\epsilon_F}\right)^2\right]
-\frac{4}{3\,Y_n}\frac{U^{(0)}_n}{\rho}\frac{5\pi^2}{12\,Y^{4/3}_n}
\left(\frac{k_BT}{\epsilon_F}\right)^2+\frac{3\,Y_e}{Y_n}\frac{k_BT}{m_n}
\nonumber\\
&-&\frac{5}{3}\frac{U^{(0)}_p}{\rho}\left[1+\frac{5\pi^2}
{12\,Y^{4/3}_p}\left(\frac{k_BT}{\epsilon_F}\right)^2\right]
-\frac{\mu_e}{m_n}Y_e-\frac{\mu_{\nu_e}}{m_n}Y_{\nu_e}+\frac{\mu_{\nu_e}}
{m_n}Y_L.  
\ee
By putting in typical values for the core ($Y_n\sim 0.7,\,Y_L\sim
0.34,\,\epsilon_F\sim 60\,{\rm MeV},\,k_BT\sim\,20\,{\rm MeV},\,
\mu_e-\mu_{\nu_e}\sim 30\,{\rm MeV}$) the Silk determinant is 
approximately given by
\be
n\,T\left(\frac{\partial s}{\partial\rho}\right)_{T,\mu_{\nu_e}}
\simeq \frac{5-10\,{\rm MeV}}{m_n},  
\label{silknumber}
\ee
which of course, is greater than zero.  In order to realistically assess this
result, we must note that the typical energy scales in this problem are 
given by the thermal energy per particle $k_BT\sim 20\,{\rm MeV}$ and
the Fermi energy $\epsilon_F\sim 60\,{\rm MeV}$.  Quantitatively, the value
of the Silk determinant given above is obtained by adding and subtracting
characteristic energies of order $\sim 20-60\,{\rm MeV}$.  Therefore, the 
answer given by eq. (\ref{silknumber}) should be taken with extreme caution
and a more careful calculation is necessarily required.    

Even though the Silk determinant
is positive, a given compressible wave is not necessarily driven.
Unstable driving from the entropy perturbation must 
overcome the loss of degeneracy pressure due to lepton diffusion 
during compression.  From the form of the pressure perturbation given 
by eq. (\ref{deltaPbig}) or equivalently from the asymptotic growth rate
given eq. (\ref{growthrate}), Silk driving or diffusive heating will occur if
\be
 \frac{\kappa_{\rm eff}}{k^2_BT}\left(\frac{\partial P}
{\partial T}\right)_{\rho,\mu_{\nu_e}}
\,n\,T\left(\frac{\partial s}
{\partial\rho}\right)_{T,\mu_{\nu_e}}+  \pi^2\kappa_0 
\left(\frac{\partial P}{\partial\mu_{\nu_e}}\right)_{\rho,
\mu_{\nu_e}}\,n\left(\frac{\partial Y_L}{\partial\rho}\right)_{T,
\mu_{\nu_e}}
>0.  
\ee
Derivatives of the total pressure with respect to temperature 
and neutrino chemical potential are approximately given by
\be
\left(\frac{\partial P}{\partial T}\right)_{\rho,\mu_{\nu_e}}=
\left(\frac{\partial P}{\partial T}\right)_{\rho,
\mu_e,\mu_{\nu_e}}+\left(\frac{\partial\mu_e}{\partial T}\right)_{\rho,
\mu_{\nu_e}}\left(\frac{\partial P}{\partial\mu_e}\right)_{\rho,T,
\mu_{\nu_e}}\simeq
\frac{\pi^2}{3}\left(\frac{k_BT}{\epsilon_F}\right)\,n\,k_B\left[
Y^{1/3}_n+Y^{1/3}_p\right]+n\,k_BY_e,
\ee
\be
\left(\frac{\partial P}{\partial\mu_{\nu_e}}\right)_{\rho,T}=
\left(\frac{\partial P}{\partial\mu_{\nu_e}}\right)_{\rho,T,\mu_e}
+\left(\frac{\partial\mu_e}{\partial\mu_{\nu_e}}\right)_{\rho,T}\left(
\frac{\partial P}{\partial\mu_e}\right)_{\rho,T,\mu_{\nu_e}}\simeq
n\,Y_{\nu_e}+n\,Y_e=n\,Y_L.  
\ee
Driving from radiative diffusion therefore occurs if
\be
\kappa_{\rm eff}\left[\frac{\pi^2}{3}\left(\frac{k_BT}{\epsilon_F}
\right)\left[Y^{1/3}_n+Y^{1/3}_p\right] +Y_e\right]\times\frac{1}{2}
\gtrsim\kappa_0\pi^2\,Y^2_L  
\ee
which is marginally satisfied for $k_BT\gtrsim 20\,\,{\rm MeV}$ and/or $\rho
\lesssim 10^{14}\,{\rm g\,cm^{-3}}$.  

\section{Appendix C: Chemical and Thermal Equilibrium}\label{a:c}

\section{ Chemical and Thermal Source Terms}

In order to calculate the conditions for chemical and thermal equilibrium, 
we must linearize the Boltzmann equation for all six ($n,p,e^-,e^+,\nu_e,
\bar{\nu_e}$) particle species.  Electron (beta) capture is the inelastic 
radiative process that mediates chemical and thermal balance i.e., 
\be
e^-+p&\leftrightarrows& n+\nu_e\\
e^++n&\leftrightarrows& p+{\bar{\nu_e}}.
\ee
In order to close our set of equations and for the sake of simplicity, 
we make five simplifying assumptions before doing so:
\begin{itemize}
\item All species are represented, to lowest order, by a Fermi-Dirac 
distribution function characterized by an individual temperature and 
chemical potential.  For example, in the case of protons
\be
f_{p}=\frac{1}{{e^{\left({E_p-\mu_p}\right)/{k_BT_p}}+1}}.
\ee
\item Thermal equilibrium for the background state,
\be
T_p=T_n=T_{e^-}=T_{e^+}=T_{{\nu_e}}=T_{\bar{\nu_e}}=T.
\ee
\item Chemical equilibrium for the background state,
\be
\frac{\mu_p+\mu_e-\mu_n}{kT_g}-\frac{\mu_{\nu_e}}{kT_{\nu_e}}.
\ee
\item Due to strong and electromagnetic interactions
\be
\delta T_{e^-}=\delta T_{e^+}=\delta T_{p}=\delta T_{n}=\delta T_g
\ee
and
\be
\delta T_{\nu_e}=\delta T_{\bar{\nu_e}}.
\ee
\item The last assumption is that the medium is
sufficiently hot and optically thick such that $E_e\simeq E_{\nu}>>
(m_n-m_p)c^2$.  
\end{itemize}

Now, consider the evolution of the partial fraction $Y_{e^-}$
mediated by the reaction $e^-+p\rightarrow n +\nu_e$, governed by
the Boltzmann equation
\be
n\left(\frac{dY_{e^-}}{dt}\right)_{{\rm capture}}
=-\frac{G^2_F}{\pi\left(\hbar c\right)^4}
\frac{\left(g^2_V+3g^2_A\right)}{\left(2\pi\right)^3}
\frac{2\pi\,c}{\left(\hbar c\right)^3}\frac{2}{h^3c^3}\int d^3p_p
\,\frac{2}{h^3c^3}\int d^3p_n\,\int^{\infty}_0\,dE_0\,E^2_0\,d\mu_0
E^2_0\,f_p\left(1-f_n\right)f_e\left(1-f_{\nu_e}\right)
\ee
where $E_0$ is the electron energy.  At the same time, we must consider
electron emission via neutrino capture via the reaction $\nu_e+n
\rightarrow e^-+p $.  This gives us
\be
n\left(\frac{dY_{e^-}}{dt}\right)_{{\rm emission}}
=+\frac{G^2_F}{\pi\left(\hbar c\right)^4}
\frac{\left(g^2_V+3g^2_A\right)}{\left(2\pi\right)^3}
\frac{2\pi\,c}{\left(\hbar c\right)^3}\frac{2}{h^3c^3}\int d^3p_p
\,\frac{2}{h^3c^3}\int d^3p_n\,\int^{\infty}_0\,dE_0\,E^2_0\,d\mu_0
E^2_0\,f_n\left(1-f_p\right)f_{\nu_e}\left(1-f_e\right).
\ee
After a quick algebraic manipulation, we may write the familiar 
expression
\be
n\frac{dY_{e^-}}{dt}& = &\left(\frac{dY_{e^-}}{dt}\right)_{{\rm capture}}
-\left(\frac{dY_{e^-}}{dt}\right)_{{\rm emission}}\nonumber\\
& = &  -\frac{G^2_F}{\pi\left(\hbar c\right)^4}
\frac{\left(g^2_V+3g^2_A\right)}{\left(2\pi\right)^3}
\frac{2\pi\,c}{\left(\hbar c\right)^3}\times\nonumber\\
&\,&\frac{2}{h^3c^3}\int d^3p_p
\,\frac{2}{h^3c^3}\int d^3p_n
\,\int^{\infty}_0\,dE_0\,d\mu_0
E^4_0\,f_p\left(1-f_n\right)f_e\left(1-f_{\nu_e}\right)\times
\left(1-e^{\frac{E_p+E_e-E_n}{kT_g}-\frac{E_{\nu_e}}{kT_{\nu_e}}+
\frac{\mu_n-\mu_p-\mu_e}{kT_g}+\frac{\mu_{\nu_e}}{kT_{\nu_e}} }\right).
\ee
Note that the term in the exponential on the right hand side is 
equal to zero in equilibrium.

Now we can consider perturbations.  By applying energy conservation 
we have
\be
n\,\delta \frac{dY_{e^-}}{dt} & = & +\frac{G^2_F}{\pi\left(\hbar c\right)^4}
\frac{\left(g^2_V+3g^2_A\right)}{\left(2\pi\right)^3}
\frac{2\pi\,c}{\left(\hbar c\right)^3}\frac{2}{h^3c^3}\int d^3p_p
\,\frac{2}{h^3c^3}\int d^3p_n
\,\int^{\infty}_0\,dE_0\,d\mu_0
E^4_0\,f_p\left(1-f_n\right)f_e\left(1-f_{\nu_e}\right)\nonumber\\
& & \times
\left[\left(\frac{E_0-\mu_{\nu_e}}{kT^2}\right)\times
\left(\delta T_g-\delta T_{\nu_e}\right)+\frac{1}{kT}\left(
\delta\mu_p+\delta\mu_e-\delta\mu_{\nu_e}-\delta\mu_n\right)\right].
\ee
If we neglect the final state blocking of the baryons, we are left
with
\be
\delta \frac{dY_{e^-}}{dt} & = & +\frac{G^2_F}{\pi\left(\hbar c\right)^4}
\frac{\left(g^2_V+3g^2_A\right)}{\left(2\pi\right)^3}
\frac{4\pi\,c Y_p}{\left(\hbar c\right)^3}
\,\int^{\infty}_0\,dE_0\,
E^4_0\,f_e\left(1-f_{\nu_e}\right)
\left[\left(\frac{E_0-\mu_{\nu_e}}{kT^2}\right)\times
\left(\delta T_g-\delta T_{\nu_e}\right)+\frac{1}{kT}\left(
\delta\mu_p+\delta\mu_e-\delta\mu_{\nu_e}-\delta\mu_n\right)\right].
\ee
This integral can easily be calculated by making use of the relation
\be
f(x_1)\left(1-f(x_2)\right)=\frac{f(x_1)-f(x_2)}{1-e^{x_1-x_2}}.
\ee
The expression for electron evolution becomes
\be
-i\omega\delta Y_{e^-}+\delta{\bf v}\cdot{\mathbf \nabla} Y_{e^-}& = &
-\frac{c\,{\tilde{\sigma}}_0}{8\pi^2}\left(\frac{kT}{m_ec^2}\right)^2
\,\left(\frac{kT}{\hbar c}\right)^3\frac{Y_p}{1-e^{y/kT}}\times
\nonumber\\
& &\left[
\left(\frac{\delta T_g-\delta T_{\nu_e}}{T}\right)\left[F_5(\eta_e)-
F_5(\eta_{\nu_e}) \right]+\left(\frac{\delta\mu_p+\delta\mu_e
-\delta\mu_n-\delta\mu_{\nu_e}}{kT}-\frac{\mu_{\nu_e}}{kT}
\frac{\delta T_g-\delta T_{\nu_e}}{T}\right)\left[F_4(\eta_e)
-F_4(\eta_{\nu_e})\right]\right]
\ee  
where
\be
{\tilde\sigma}_0\equiv\sigma_0\left(g^2_V+3g^2_A\right)\nonumber\\
G^2_F=\frac{\pi\sigma_0}{4}\frac{\left(\hbar c\right)^4}{\left(
m_ec^2\right)^2}\nonumber\\
y\equiv \mu_{\nu_e}-\mu_e\nonumber\\
F_n(\eta)=\int^{\infty}_0dx\,\frac{x^n}{e^{x-\eta}+1}.
\ee
If we wish to follow the evolution of the total electron fraction $Y_e=
Y_{e^-}-Y_{e^+}$ where the evolution of the $e^+s$ are  mediated by the 
reactions $e^++n\rightleftarrows p+{\bar{\nu_e}}$, then the relevant 
expression becomes
\be
-i\omega\delta Y_{e}+\delta{\bf v}\cdot{\mathbf \nabla} Y_{e}& = &
-\frac{c\,{\tilde{\sigma}}_0}{8\pi^2}\left(\frac{kT}{m_ec^2}\right)^2
\,\left(\frac{kT}{\hbar c}\right)^3\frac{Y_p}{1-e^{y/kT}}\times
\nonumber\\
& &\left[
\left(\frac{\delta T_g-\delta T_{\nu_e}}{T}\right)\left[D_5(\eta_e)-
D_5(\eta_{\nu_e}) \right]+\left(\frac{\delta\mu_p+\delta\mu_e
-\delta\mu_n-\delta\mu_{\nu_e}}{kT}-\frac{\mu_{\nu_e}}{kT}
\frac{\delta T_g-\delta T_{\nu_e}}{T}\right)\left[D_4(\eta_e)
-D_4(\eta_{\nu_e})\right]\right]
\label{deltaYe}
\ee
and the expression for conservation of electron neutrino fraction, which is 
defined as $Y_{\nu_e}\equiv Y_{\nu_e}-Y_{\bar{\nu_e}}$, reads
\be
-i\omega\delta Y_{\nu_e}+\delta{\bf v}\cdot{\mathbf \nabla} Y_{\nu_e}& = &
+\frac{c\,{\tilde{\sigma}}_0}{8\pi^2}\left(\frac{kT}{m_ec^2}\right)^2
\,\left(\frac{kT}{\hbar c}\right)^3\frac{Y_p}{1-e^{y/kT}}\times
\Biggl\{
\left(\frac{\delta T_g-\delta T_{\nu_e}}{T}\right)\left[D_5(\eta_e)-
D_5(\eta_{\nu_e}) \right]\nonumber\\&+&\left(\frac{\delta\mu_p+\delta\mu_e
-\delta\mu_n-\delta\mu_{\nu_e}}{kT}-\frac{\mu_{\nu_e}}{kT}
\frac{\delta T_g-\delta T_{\nu_e}}{T}\right)\left[D_4(\eta_e)
-D_4(\eta_{\nu_e})\right]\Biggr\}-\frac{i}{n_B}{\bf k}\cdot\delta
{\bf F}_L
\ee
where we have defined
\be
D_n\equiv F(\eta)-F(-\eta).
\ee

The energy source term for the electrons is calculated in a very similar 
manner.
To be clear, the first law of thermodynamics for the electrons may be written
as
\be
\frac{d U_{e^-}}{dt}+\left(U_{e^-}+P_{e^-}\right)
{\mathbf\nabla}\cdot{\bf v}& = &S_{e^-}
=   -\frac{G^2_F}{\pi\left(\hbar c\right)^4}
\frac{\left(g^2_V+3g^2_A\right)}{\left(2\pi\right)^3}
\frac{2\pi\,c}{\left(\hbar c\right)^3}\times\\
&\, &\frac{2}{h^3c^3}\int d^3p_p
\,\frac{2}{h^3c^3}\int d^3p_n
\,\int^{\infty}_0\,dE_0\,d\mu_0
E^5_0\,f_p\left(1-f_n\right)f_e\left(1-f_{\nu_e}\right)\times
\left(1-e^{\frac{E_p+E_e-E_n}{kT_g}-\frac{E_{\nu_e}}{kT_{\nu_e}}+
\frac{\mu_n-\mu_p-\mu_e}{kT_g}+\frac{\mu_{\nu_e}}{kT_{\nu_e}} }\right).
\nonumber
\ee
Upon perturbations, conservation of energy per unit volume for the 
the electrons is expressed by 
\be
-i\omega\delta U_{e^-}+\delta{\bf v}\cdot{\mathbf \nabla} U_{e^-}
+\left(U_{e^-}+P_{e^-}\right){\mathbf\nabla}\cdot\delta{\bf v}& = &
-\frac{c\,{\tilde{\sigma}}_0}{8\pi^2}\left(\frac{kT}{m_ec^2}\right)^2
\,\frac{\left(kT\right)^4}{\left(\hbar c\right)^3}\frac{n_p}{1-e^{y/kT}}
\times
\Biggl\{
\left(\frac{\delta T_g-\delta T_{\nu_e}}{T}\right)\left[F_6(\eta_e)-
F_6(\eta_{\nu_e}) \right]\nonumber\\
&+&\left(\frac{\delta\mu_p+\delta\mu_e
-\delta\mu_n-\delta\mu_{\nu_e}}{kT}-\frac{\mu_{\nu_e}}{kT}
\frac{\delta T_g-\delta T_{\nu_e}}{T}\right)\left[F_5(\eta_e)
-F_5(\eta_{\nu_e})\right]\Biggr\}.
\ee  
Note that the source term which drives electrons into thermal equilibrium
is {\it not exactly equal} to the source term which drives the electrons
into chemical equilibrium even though they are comparable in both appearance
and magnitude.  This is of importance when considering a spectrum of 
perturbations in the upper layer of a proto-neutron star.  For a given 
dynamical timescale, the depth beneath the neutrinosphere at which the the 
fluctuations are in chemical equilibrium may differ, in principle, from 
the layer in which thermal equilibrium is achieved.  In terms of neutrino 
bubble
driving, whether or not the perturbations are either in chemical or thermal 
equilibrium or both, will determine whether or not driving due to 
changes in neutrino chemical potential (or the lepton flux) or driving 
due to changes in neutrino temperature (or the radiative flux) dominate the 
mechanics of the instability.  

The first law for the combined electron, positron, neutron, and proton gas
is given by
\be
-i\omega\delta U_g+\delta{\bf v}\cdot{\mathbf \nabla} U_g
+\left(U_g+P_g\right){\mathbf\nabla}\cdot\delta{\bf v}& = &
-\frac{c\,{\tilde{\sigma}}_0}{8\pi^2}\left(\frac{kT}{m_ec^2}\right)^2
\,\frac{\left(kT\right)^4}{\left(\hbar c\right)^3}\frac{n_p}{1-e^{y/kT}}
\times
\Biggl\{
\left(\frac{\delta T_g-\delta T_{\nu_e}}{T}\right)\left[C_6(\eta_e)-
C_6(\eta_{\nu_e}) \right]\nonumber\\
&+&\left(\frac{\delta\mu_p+\delta\mu_e
-\delta\mu_n-\delta\mu_{\nu_e}}{kT}-\frac{\mu_{\nu_e}}{kT}
\frac{\delta T_g-\delta T_{\nu_e}}{T}\right)\left[C_5(\eta_e)
-C_5(\eta_{\nu_e})\right]\Biggr\}
\label{deltaUgas}
\ee
where 
\be
U_g\equiv U_n+U_p+U_e\,\,\,\,\,\,&{\rm and}&
\,\,\,\,\,\,U_e\equiv U_{e^-}+U_{e^+}\nonumber\\
P_g\equiv P_n+P_p+P_e\,\,\,\,\,\,&{\rm and}&
\,\,\,\,\,\,P_e\equiv P_{e^-}+P_{e^+}\nonumber\\
C_n(\eta)&\equiv& F_n(\eta)+F_n(-\eta).
\ee
For the electron type neutrino's, the first law of thermodynamics reads
\be
-i\omega\delta U_{\nu_e}+\delta{\bf v}\cdot{\mathbf \nabla} U_{\nu_e}
+\left(U_{\nu_e}+P_{\nu_e}\right){\mathbf\nabla}\cdot\delta{\bf v}& = &
+\frac{c\,{\tilde{\sigma}}_0}{8\pi^2}\left(\frac{kT}{m_ec^2}\right)^2
\,\frac{\left(kT\right)^4}{\left(\hbar c\right)^3}\frac{n_p}{1-e^{y/kT}}
\times
\Biggl\{
\left(\frac{\delta T_g-\delta T_{\nu_e}}{T}\right)\left[C_6(\eta_e)-
C_6(\eta_{\nu_e}) \right]\nonumber\\
&+&\left(\frac{\delta\mu_p+\delta\mu_e
-\delta\mu_n-\delta\mu_{\nu_e}}{kT}-\frac{\mu_{\nu_e}}{kT}
\frac{\delta T_g-\delta T_{\nu_e}}{T}\right)\left[C_5(\eta_e)
-C_5(\eta_{\nu_e})\right]\Biggr\}-i\,{\bf k}\cdot\delta{\bf F}
\ee
where now $U_{\nu_e}\equiv U_{\nu_e}+U_{{\bar{\nu_e}}}$ and 
$P_{\nu_e}\equiv P_{\nu_e}+P_{{\bar{\nu_e}}}$.  Notice that we have not 
included 
the non-electronic type neutrinos in this discussion.  The only 
mechanism which thermodynamically couples the $\mu$- and $\tau$-type 
neutrinos to the other fluid and radiation components is  
inelastic scattering with electrons, in other words, the inverse 
Compton effect.  This coupling is heavily blocked by the 
final state occupancy of the electrons which reduces the 
scattering cross-section by a factor of $(1-f_e)$.

\section{Characteristic Timescales for Chemical and Thermal Equilibrium}

Timescales for 
both chemical and thermal equilibrium dictate whether or not 
the source terms represented in eqs. (\ref{deltaYe}) and (\ref{deltaUgas}) 
vanish.  That is, electron capture and emission dictate {\it both} 
chemical and thermal equilibrium simultaneously.

In order to calculate the magnitude of the lepton source term in eq.
(\ref{deltaYe}), it is convenient to write the left hand side as
\be
-i\omega\delta Y_e+\delta{\bf v}\cdot{\mathbf\nabla} Y_e
=-i\omega\left[\left(\frac{\partial Y_e}{\partial T_g}\right)_{\eta_e,\rho}
\delta T_g+\left(\frac{\partial Y_e}{\partial \eta_e}\right)_{T_g,\rho}
\delta\eta_e+ \left(\frac{\partial Y_e}{\partial\rho}\right)_{T_g,\eta_e}
\delta\rho   \right]+\delta{\bf v}\cdot{\mathbf\nabla}Y_e.
\ee
Thus, the source of electrons will vanish if 
\be
\omega_{Y,{\rm th}}\equiv \frac{c\sigma_a n_0Y_p}{T\left({\partial Y_e}
/{\partial T_g}\right)_{\eta_e,\rho}}\times\frac{\left[D_5(\eta_e)-
D_5(\eta_{\nu_e})\right]}{1-e^{y/kT}}\gg\omega
\label{rate1}
\ee
and
\be
\omega_{Y,{\rm chem}}\equiv \frac{c\sigma_a\,n_0Y_p}{(\partial Y_e/
\partial\eta_e)_{T_g,\rho}}\times\frac{\left[D_4(\eta_e)-D_4(\eta_{\nu_e})
\right]}{1-e^{y/kT}}\gg\omega
\label{rate2}
\ee
where we have defined
\be
\sigma_a&\equiv&\frac{{\tilde\sigma_0}}{8\pi^2}\left(\frac{kT}{m_ec^2}\right)^2
\nonumber\\
n_0&\equiv & \left(\frac{k_BT}{\hbar\,c}\right)^3
\ee
where $\sigma_a$ is the characteristic absorption opacity and $n_0$ is a
characteristic relativistic particle number density.  
Note that the expression for 
$\omega_{Y,{\rm th}}$ was obtained by rewriting the condition for chemical 
equilibrium as
\be
\frac{\delta\mu_p+\delta\mu_e-\delta\mu_n+\delta\mu_{\nu_e}}{kT}-
\frac{\mu_{\nu_e}}{kT}\frac{\delta T_g-\delta T_{\nu_e}}{T}=
\delta\eta_p+\delta\eta_e-\delta\eta_n-\delta\eta_{\nu_e}.  
\label{balance}
\ee  
Likewise, in order to calculate the net effect of the energy source term 
in eq. (\ref{deltaUgas}), it is useful to write the left hand side as
\be
-i\delta U_g+\delta{\bf v}\cdot {\mathbf\nabla} U_g+
i(U_g+P_g){\bf k}\cdot\delta{\bf v}=
-i\left[\left(\frac{\partial U_g}{\partial T_g}\right)_{\eta_e,\rho}
\delta T_g
+\left(\frac{\partial U_g}{\partial\eta_e}\right)_{T_g,\rho}
\delta\eta_e+ \left(\frac{\partial U_g}{\partial\rho}\right)_{T_g,
\eta_e}\delta\rho\right]+\delta{\bf v}\cdot {\mathbf\nabla} U_g+
i\,(U_g+P_g)\,{\bf k}\cdot\delta{\bf v}.  
\ee
The heat source 
for the {\it npe$\gamma$} gas will vanish if
\be
\omega_{U,{\rm th}}\equiv \frac{c\,\sigma_a\, n\,Y_p\, U_0}{T(\partial U_g/
\partial T_g)_{\eta_e,\rho}}\times\frac{\left[C_6(\eta_e)-
C_6(\eta_{\nu_e})\right]}{1-e^{y/kT}}\gg\omega
\label{rate3}
\ee   
and
\be
\omega_{U, {\rm chem}}\equiv\frac{c\,\sigma_a\,n\,Y_p\,U_0}
{(\partial U_g/\partial\eta_e)_{T_g,\rho}}\times
\frac{\left[C_5(\eta_e)-
C_5(\eta_{\nu_e})\right]}{1-e^{y/kT}}\gg\omega
\label{rate4}
\ee
where
\be
U_0\equiv \frac{k_B^4T^4}{\hbar^3c^3}
\ee
serves as a characteristic energy density for relativistic particles.  The 
equilibrition rate $\omega_{Y,{\rm chem}}$, given by eq. (\ref{rate2}), 
is identical to that derived by Burrows \& Lattimer (1986, their eq. 16) 
except for the presence of a factor 
$(\partial Y_e/\partial\eta_e)_{T_g,\rho}$ in the 
denominator and the absence of a term that drives the 
system into statistical equilibrium in the numerator.  Both discrepencies
can be readily accounted for by realizing that our analysis is concerned
with perturbations
about an equilibrium already in statistical and thermal balance, while 
those authors calculate whether or not their equilibrium is in 
statistical and thermal balance with respect to the contraction timescale.

In what follows, we approximate whether or not acoustic-type 
fluctuations near the neutrinosphere are in chemical and thermal 
equilibrium.  We conclude that chemical and thermal balance 
between the radiation and gas is marginally satisfied 
at the neutrinosphere.

\section{Requirements for Chemical and Thermal Equilibrium Near the 
Neutrinosphere}

Interestingly, the chemical and energy source terms 
which lead to the characteristic frequencies $\omega_{Y,{\rm th}}$ and
$\omega_{\rm U, th}$ vanish.  This results from the mathematical fact 
that $F_n(\eta)$ is an even function if $n$ is odd and an odd function
if $n$ is even.  From a more physical point of view, chemical and 
thermal balance {\it simultaneously occur} as long as eq. (\ref{balance})
is equal to zero. 

Throughout the star, from the core to the neutrinosphere, the electrons 
can be considered fully degenerate such that $\eta_e>>1$.  The thermal 
contribution to the electrons is more profound near the neutrinosphere, but 
there, $\eta_e$ is still $\gtrsim$ a few.  Thus, the Fermi-Dirac 
integrals for the electrons $F_n(\eta_e)$ can be approximated as 
(Pathria 1996)
\be
F_n\left(\eta_e\right)\simeq\frac{\Gamma(n+1)}{\Gamma(n+2)}\,\eta^{n+1}_e
=\frac{\eta^{n+1}_e}{n+1},
\ee
especially if $n$ is large.  Note that $\Gamma(n)$ is the Euler Gamma 
function.  Also, near the neutrinosphere, the chemical 
potential of the electron-type neutrinos $\mu_{\nu_e}\sim$ a few times 
smaller than $\mu_e$.   As a result, we may write 
\be
\omega_{Y,{\rm chem}}&\simeq& \frac{2}{5}\times\frac{c\sigma_a\,n_0Y_p}
{(\partial Y_e/\partial\eta_e)_{T_g,\rho}}\,{\eta^5_e}\nonumber\\
{\rm and}\,\,\,\,\omega_{U,{\rm chem}}&\simeq&\frac{1}{3}\times
\frac{c\,\sigma_a\, n\,Y_p\, U_0}{T(\partial U_g/
\partial T_g)_{\eta_e,\rho}}\eta^6_e.
\ee
Putting in numerical values typical for conditions at and around the 
neutrinosphere allows us to write the characteristic frequencies for 
chemical and thermal equilibrium in terms of $Y_e,\,T,\,$and $\mu_e$,
\be
\omega_{Y,{\rm chem}}&\simeq& 3\times 10^4\,\frac{\mu^6_{20}}{T_4}\,{\rm 
s}^{-1}
\nonumber\\
\omega_{U,{\rm chem}}&\simeq& 2\times 10^4\,Y_{1/3}\frac{\mu^6_{20}}
{T_4}\,{\rm s}^{-1}
\ee
where $Y_{1/3}$, $T_4$, and $\mu_{20}$ are the electron fraction in units 
of $1/3$, temperature in units 4 MeV, and electron chemical potential 
normalized to 20 MeV, respectively.  In the above formulae, we made the 
approximation that the gas is primarily supported by the thermal 
pressure of the baryons.  At the neutrinosphere, the sound crossing time
$\sim 0.1$ ms, implying that chemical and thermal 
equilibrium is marginally satisfied for acoustic-type oscillations.
\footnote{If the atmosphere is dominated by magnetic pressure, the 
acoustic time for the fast magnetosonic wave is shorter than that of 
ordinary sound waves.  Even if this were the case, the slow magnetosonic
wave is the excitation species of interest since it always has 
larger growth rates than the fast wave (BS03).}

\end{document}